\documentclass{hep99}
\usepackage{epsf}
\newcommand{\lsim}{\,{\buildrel < \over {_\sim}}\,}
\newcommand{\gsim}{\,{\buildrel > \over {_\sim}}\,}
\begin{document}

\title{High Energy Nuclear Collisions}

\author{K. J. Eskola \\
\vspace{-5cm}
\begin{flushright}
{\normalsize JYFL-10/99 \\
hep-ph/9911350\\
}
\vspace{4cm}
\end{flushright}
}

\address{Department of Physics, University of Jyv\"askyl\"a,
P.O. Box 35, FIN-40351 Jyv\"askyl\"a, Finland\\[3pt] 
E-mail: {\tt kari.eskola@phys.jyu.fi}}
 
\abstract{ Highlights of the results from ultrarelativistic heavy ion
collisions at CERN-SPS are reviewed.  In particular, I discuss how the
experimental results indicate that a collective strongly interacting
system has been produced, and what are the implications towards the
Quark Gluon Plasma.  The physical ideas behind measuring certain
observables are introduced. The future program of high energy nuclear
collisions at BNL-RHIC and CERN-LHC/ALICE is also briefly discussed.
\\ \\ 
{\small Plenary talk at the International Europhysics Conference
on High Energy Physics, EPS-HEP99, Tampere, Finland, July 1999.  } }
\maketitle
 
\section{Introduction} 

This talk is organized as follows: In the first, introductory section,
I will briefly recapitulate the physics motivations for colliding
heavy ions at ultrarelativistic energies. For the second section, I
have selected some of the experimental highlights from
ultrarelativistic heavy ion collisions (let me use an acronym URHIC)
at CERN-SPS. For each topic, the basic idea will be explained first,
followed then by a discussion of the interesting experimental
observations.  As URHIC at the SPS represent the highest energy
nuclear collisions so far, they and their implications will have the
main emphasis in this review. In the third section, I will briefly
discuss the future program for heavy ions, BNL-RHIC and the ALICE
experiment at CERN-LHC.  In the fourth section, I will present my
conclusions.  For recent reviews on the same subject, see e.g.
\cite{HEINZsew,HEINZqm99,BLAIZOTqm99,BASS}.

\subsection{General facts} 

The main goal of ultrarelativistic heavy ion collisions is to
study the thermodynamics of strongly interacting matter. In colliding
two heavy nuclei ($A\sim 200$) together at very high cms energies
($\sqrt s \sim 20...200...5500\,A$GeV), the primary aim is to
produce experimentally a new phase of matter, the Quark-Gluon Plasma
(QGP), and -- even more importantly -- to {\em observe} the QCD phase
transition from the QGP phase to the Hadron Gas (HG)
phase. Cosmological motivation for URHIC is the fact that our
Universe has undergone such a transition within its first microseconds.

In general, although the basic motivation for such studies comes from
the theory, it is fair to state that the field of URHIC is an
experimentally driven one. In this field, it is very difficult to 
perform calculations from first principles, so -- in addition to
precision measurements --  good, QCD-based phenomenology is needed.

In relation to conventional high energy particle physics
(HEPP), URHIC naturally share the goal of making high precision
measurements.  However, the two fields are orthogonal with respect to
their physics goals: where HEPP aims for detection of new particles
and new symmetries (Higgs, SuSy), URHIC reach for detection of a new
phase of matter, made of the known QCD-quanta.  In HEPP, simplicity is
obtained through producing as {\em few} particles in the final state
as possible, whereas in URHIC simplicity is obtained by producing as
{\em many} particles in the final state as possible: only if the
strongly interacting system becomes dense and large enough it can become
collective and laws of thermodynamics become applicable. In other
words, for URHIC the goal is to do elementary particle condensed
matter physics.  In this, and throughout my talk as well, the keywords
are {\em lifetime} and {\em volume} of the produced, extended, system.

\subsection{QCD phase diagrams}

\begin{figure}[htb] 
\begin{center} 
\vspace*{1.5cm} 
\centerline{\hspace{-4.cm}\epsfxsize=5cm\epsfbox{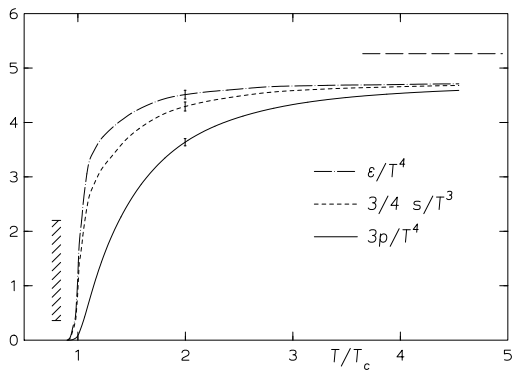}}
\end{center} 
\vspace*{-2cm} 
\caption{Scaled energy density, entropy density and pressure
vs. scaled temperature at the continuum limit in pure gauge $SU(3)$
theory \cite{BOYD,LAERMANN}. The horizontal dashed line is the ideal
gas limit and the vertical hatched band illustrates the latent
heat. The figure is from \cite{BOYD}.  }
\label{lQCD}
\end{figure}

The theoretical foundation for searching the new QCD-phase of matter
is given by the first principles calculations of lattice-QCD (for a
review, see \cite{LAERMANN}). From there we know that the QGP phase
with partonic degrees of freedom undoubtedly exists. An example of
these calculations for the Equation of State (EoS, pressure as a
function of energy density) of the QCD-matter is given in
Fig.~\ref{lQCD}. The abrupt change in the energy density from the
confined phase at low $T$ to the deconfined phase at high $T$ is a
consequence of a first order phase transition in the pure gauge
$SU(3)$.  With dynamical fermions but still with zero baryochemical
potential $\mu_B=0$ (which corresponds to zero net baryon number), the
order of the phase transition has been seen to change with the number
of quark flavours and with masses of the three lightest quarks. For
the moment, inclusion of dynamical fermions can only be done in
certain approximations and keeping $\mu_B=0$. Calculations with
$\mu_B\ne0$ are not yet available, due to the complex valued actions.

The critical deconfinement temperature $T_c$ at $\mu_B=0$ is about 265
MeV for the pure $SU(3)$ glue, with quarks the estimates lie in the
region $T_c=140...200$ MeV \cite{LAERMANN}. While the deconfinement
phase transition is studied through Polyakov loops, the phase
transition related to chiral symmetry can be studied trough chiral
condensates $\langle\bar\Psi\Psi\rangle$. In the lattice QCD
calculations, chiral symmetry has been observed to be restored
(i.e. $\langle\bar\Psi\Psi\rangle\rightarrow0$) always at the same
$T_c$ as the deconfinement transition occurs \cite{LAERMANN}.

\begin{figure}[htb]
\vspace{2.5cm}
\centerline{\hspace{-4cm}\epsfxsize=7cm\epsfbox{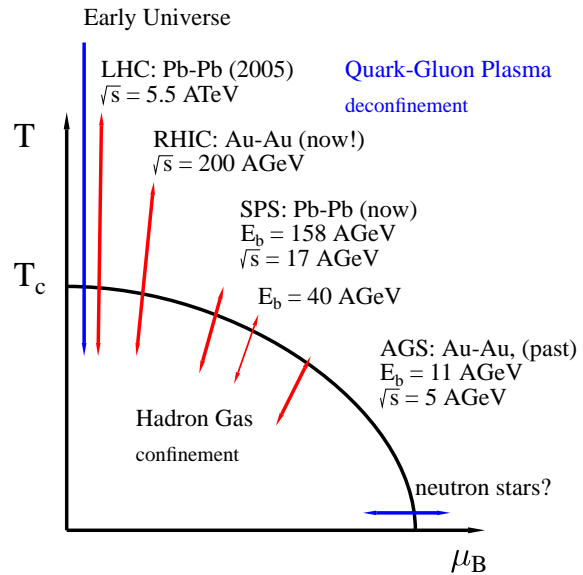}}
\vspace*{-2cm} 
\caption{ An illustration of the conventional QCD phase diagram in the
plane of chemical potential $\mu_B$ and temperature $T$. The
regions of the phase diagram the highest energy heavy ion collisions
are probing are sketched by the arrows.
}
\label{QCDpd}
\end{figure}

With URHIC, one explores the QCD phase diagram in the $\mu_B,T$-plane
as illustrated in Fig.~\ref{QCDpd}. With increasing nuclear mass
number $A$ and with increasing cms-energy, more partons are liberated
in the collision and the energy density of the initial QGP-system
(immediately after the $AA$-collision) increases.  With increasing
initial energy densities and increasing initial temperatures, sketched
by the arrows in Fig.~\ref{QCDpd}, the lifetime and the volume of the
plasma evidently increase as well, improving the chances for observing
signals directly from the QGP. Baryon stopping has already been
observed to decrease with the cms-energies \cite{ROLANDNA49}, and this
trend is expected to continue at larger $\sqrt s$.  With less net
baryon number stopped in the central rapidity region, which is the
main region of interest for the QGP studies, also $\mu_B$ will
decrease, as illustrated in the figure.  Ideally, it would of course
be beneficial to get as close as possible to the theoretically best
understood limit $\mu_B/T\rightarrow 0$.  Cosmologically, the interest
lies also in this region: in the early Universe, the inverse of the
specific entropy is tiny, $\sim 10^{-9}$.

At very high cms-energies, the formed QGP can be said to be produced
by ``heating'': the QGP consists of gluons and quark-antiquark
pairs, and the energy for creating them is provided by the high cms-energy.
The QGP can exist also at $T=0$ at large values of $\mu_B$.  This would
correspond to preparing the QGP by compressing the normal nuclear
matter beyond the critical density, which is of the order of $\sim 10$ times
the normal nuclear matter density 0.17 fm$^{-3}$. This may happen in
the cores of neutron stars but as this region is not relevant for URHIC,
I will not discuss it here. 

The current and future experimental program of highest energy nuclear
collisions is also illustrated in Fig.~\ref{QCDpd}. In addition to the
experiments I will discuss below, also an extensive intermediate and
lower energy program for heavy ions (BNL-AGS, GSI-SIS, etc.) exists
but in this talk I will concentrate only on the highest energy
collisions. In the fixed target experiments at the Super Proton
Synchrotron (SPS) of CERN, the highest beam energies have been
$E_b=200\,A$GeV for $^{32}$S, and currently $E_b=158\,A$GeV for the
Pb-beam in $^{208}$Pb+$^{208}$Pb collisions. The corresponding
nucleon-nucleon cms-energies are $\sqrt s= 20$ GeV and 17.3 GeV.  A
lower energy run at the SPS with $E_b=40\,A$GeV will take place this
year. In the near future, starting in December 1999, in the
Relativistic Heavy Ion Collider (RHIC) at the Brookhaven National
Laboratory (BNL), the collisions of highest energy and heaviest nuclei
will be $^{197}$Au+$^{197}$Au at $\sqrt s= 200\,A$GeV. In year 2005,
in A Large Ion Collider Experiment (ALICE) at the Large Hadron
Collider (LHC) at CERN, one will collide $^{208}$Pb nuclei on
$^{208}$Pb at $\sqrt s=5500\,A$GeV.

As it is expected that baryon stopping nearly vanishes at the LHC
energies, the experimental URHIC program will probe an extensive range
of the conventional QCD phase diagram in the $\mu_B,T$-plane. The
critical question is whether the QCD phase transition can indeed be
observed. Regarding the QCD phase diagram itself, there have been very
interesting theoretical developments recently: at high densities,
quarks may form Cooper pairs and a new, color superconducting, phase
may exist \cite{CSCold,CSCnew,RAJAGOPAL}. Most probably this happens
at larger values of $\mu_B$ than will be reachable experimentally by
URHIC.  On the other hand, the $\mu_B,T$-phase diagram may have a
critical or tricritical point somewhere along the phase transition
line in Fig.~\ref{QCDpd}, depending on the order of the phase
transition in full QCD -- which depends on the masses of dynamical
quarks and the effects of finite $\mu_B$. By varying the cms energy
it may be possible to trace this point down. For a recent review of
these exciting theoretical developments, see \cite{RAJAGOPAL}.

\subsection{Space-time evolution: lifetime and volume}

The basic difference between URHIC and collisions of point-like
(e$^+$+e$^-$) or small composite particles (p+$\bar {\rm p}$, p+p) is
that an ideal system produced in URHIC is an extended and collective
one with a large volume and a long lifetime. Once the system is
produced, its space-time evolution cannot be controlled.  In fact, the
only experimentally controllable initial parameters are the mass
numbers of the colliding nuclei and the collision energy. In addition,
a handle on the impact parameter in each collision can be obtained by
forming event classes of different multiplicities and transverse
energies with a correlation to the energies observed in the
zero-degree calorimeter. With these few controllable initial
parameters, information of the whole space-time evolution of the
system must be extracted from the various observables measured in the
final state.

\begin{figure}[htb]
\vspace{2.cm}
\centerline{\hspace*{1cm} \epsfxsize=9cm\epsfbox{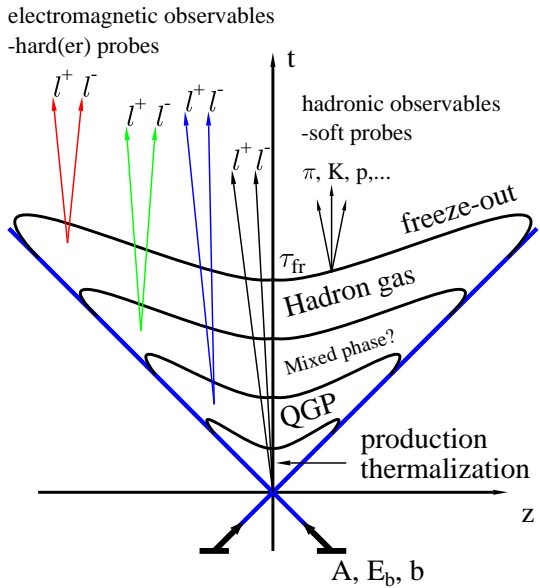}}
\vspace*{-1cm} 
\caption{ Space-time evolution of a strongly interacting system
produced in a very high energy $A+A$ collision, projected in the plane
of the longitudinal coordinate $z$ and time $t$. Different stages of
evolution and different types of observables are demonstrated. }
\label{spacetime}
\end{figure}

The different stages in the space-time evolution of a strongly
interacting system produced in an URHIC can be pictured as in
Fig.~\ref{spacetime}. At sufficiently high energies, the colliding
nuclei are Lorentz-contracted thin disks surrounded by virtual clouds
of partons. The original impact of the nuclear disks takes place
almost instantaneously (within a transit time $\tau_{\rm
tr}=2R_A/\gamma$) in a region around $z,t\sim 0$. Primary
production of gluons, quarks and antiquarks occurs in the central
rapidity region during some typical formation time $\tau_0$ of the
order of a fraction of fm$/c$, and $\tau_{\rm tr}\ll\tau_0$.  In
a dense enough system the mean free paths of the QCD-quanta are much
less than any typical homogeneity size $V$ in the system:
$\lambda_{\rm mfp}^3 \ll V$. A dense enough system reaches quickly
a {\em kinetic} equilibrium with {\em locally} thermal quark and gluon
momentum distributions.  {\em Chemical} equilibration is preferable but
not necessary to define a QGP, i.e. the density of quarks and
antiquarks relative to that of gluons does not need to coincide with
that in an ideal gas.  The system will further evolve collectively as
a QGP if it is interacting strongly enough and if the scattering times
of quarks and gluons are much less than the inverse of the expansion
rate (Hubble constant$^{-1}$) of the system. For a recent discussion, 
see e.g. \cite{HEINZsew}.

In the most ideal case -- in the limit of a very dense system -- the
evolution of the system after its formation can be described in terms
of hydrodynamics.  If the system were exactly longitudinally boost
invariant, physics would be invariant along hyperbolas of constant
longitudinal proper time $\tau=\sqrt{t^2-z^2}$ \cite{BJORKEN}. In
reality, however, already due to its edges the system is never
perfectly boost invariant, which is illustrated in
Fig.~\ref{spacetime} by the curves bending backwards in time near the
light cones. Notice also that depending on baryon stopping the
temperatures and chemical potentials in the central rapidity region
can be quite different from the ones at large rapidities.

Analogously to the early Universe in the Big Bang cosmology, the
produced QGP system expands and cools. Eventually the system ends up
in a Hadron Gas phase. If the QCD-phase transition is of first order,
a stage of a mixed phase with both QGP and HG coexisting at a (local)
critical temperature and chemical potential will appear in between.
In the end of the HG phase, the average mean free paths of hadrons
exceed the homogeneity size of the system, $\lambda_{\rm mfp}^3 > V$,
and the Hadron Gas decouples, i.e. freezes out. The hydrodynamical,
locally thermal, picture ceases to apply at this point. In terms of
hydrodynamics, the final state hadrons are emitted from the freeze-out
surface. Notice, however, that the decoupling does not not take place
instantaneously: as illustrated in Fig.~\ref{spacetime}, due to
Lorentz-dilation effects, the regions near light cones freeze out
typically later than the central regions. Furthermore, even within a
certain spatial region decoupling is never an instantaneous process
(as drawn in the figure for simplicity) but a dynamical one,
determined by the expansion rate and the hadronic rescattering cross
sections.

The probes of strongly interacting matter can be divided into hadronic
and electromagnetic ones. Typically, as illustrated in
Fig.~\ref{spacetime} the hadronic probes reflect the conditions at the
freeze-out hypersurface, i.e. at the very last stage of the collective
evolution. The electromagnetic probes, thermal lepton pairs and
photons \cite{EMPROBES,PVRqm91,WAMBACH97}, in turn are emitted
throughout the whole space-time history of the system; they decouple
immediately due to their long mean free paths.  The thermal
electromagnetic probes have to compete with the primary production
processes, e.g. Drell-Yan process for dileptons and prompt photon
production for photons, which have the same origin in hard scatterings
as in hadron-hadron collisions. Also the hadrons emitted from the
freeze-out surface may decay electromagnetically and contribute
strongly to the background for the electromagnetic probes. Because of
the rapidly falling exponential momentum distributions in the thermal
system, the momentum or mass scale of an electromagnetic probe is
typically close to the local temperature at the time of emission.
Therefore, the harder the probes are, the hotter and earlier system
they typically probe. The probes coming directly from hard scatterings
in the primary nuclear collision are the earliest ones and usually
referred to as ``hard probes'' \cite{HPC}. Some hard probes, like jet
production (also in connection with direct photons) are strongly
interacting but -- thanks to the large momentum or mass scale $Q\gg T$
involved -- they are not absorbed in the QGP or HG.

\begin{figure}[htb]
\vspace{0.5cm}
\centerline{\hspace*{-0.5cm} \epsfxsize=4cm \epsfbox{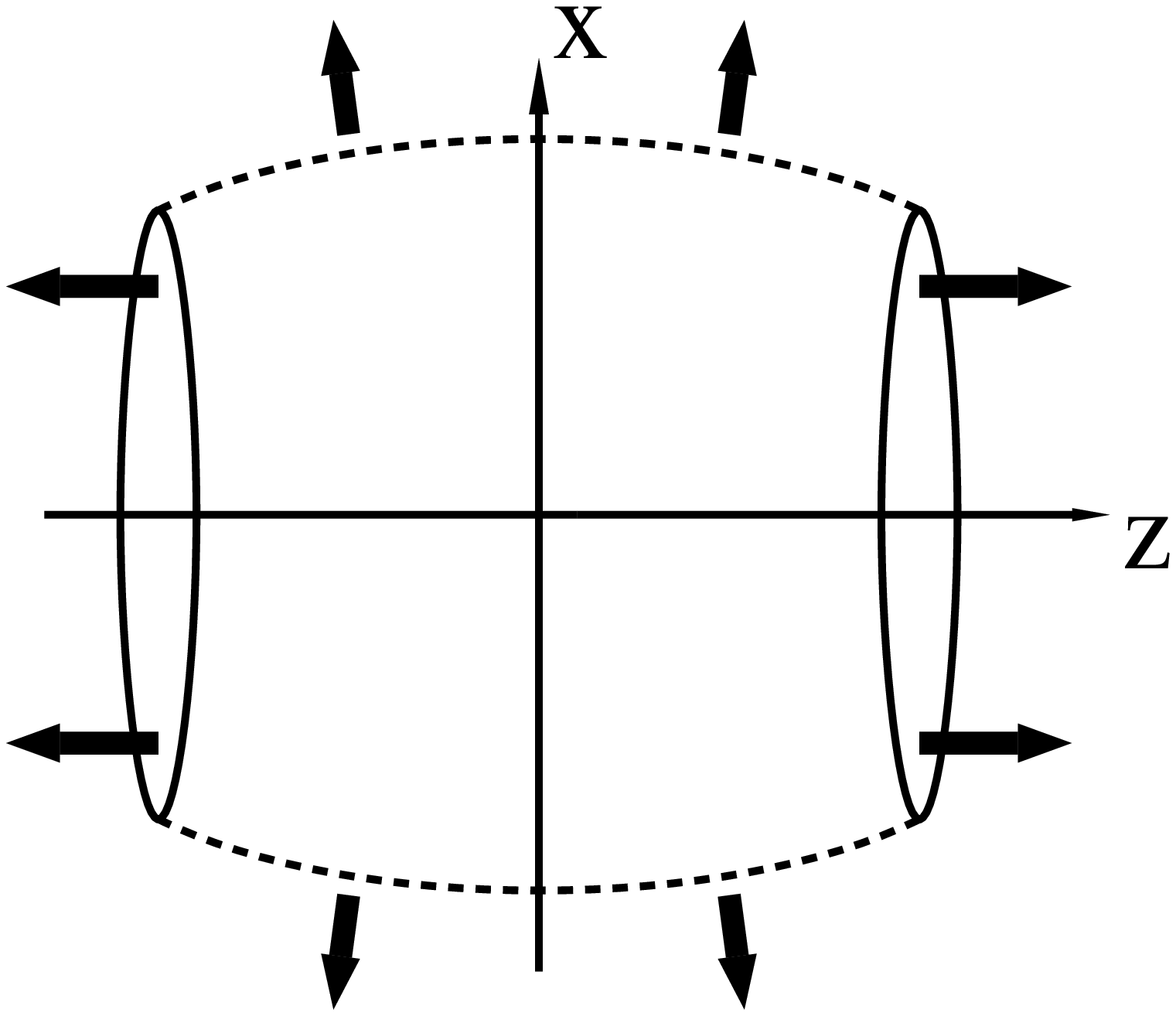}\hfill}
\vspace*{-4.2cm}
\centerline{\hspace*{3cm}\epsfxsize=4.5cm   \epsfbox{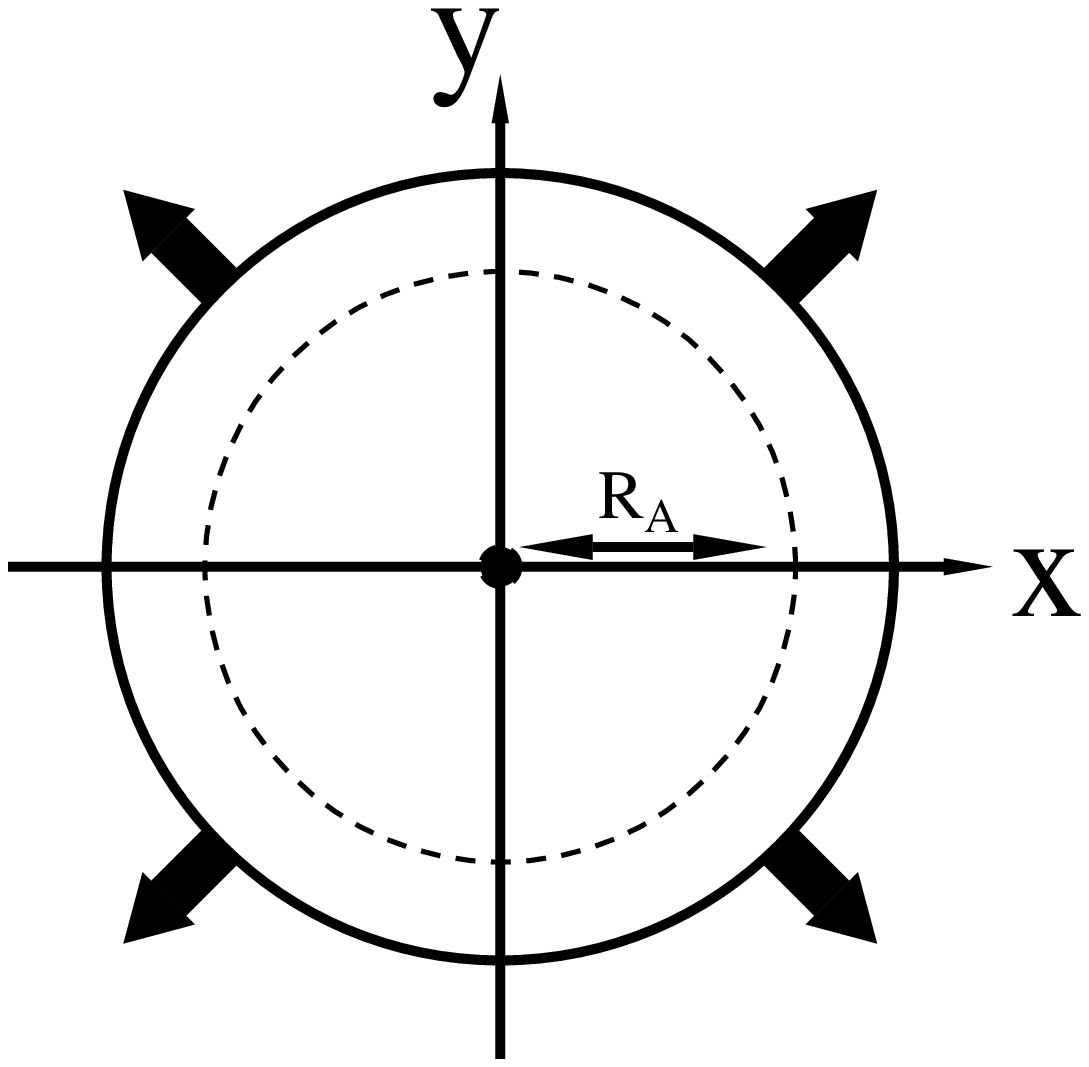} }
\vspace{-1cm}
\caption{Snapshots of an expanding system in longitudinal-transverse
coordinate $z,x$-plane, and in transverse coordinate $x,y$-plane. The
black arrows indicate the longitudinal and transverse flow. The rest frame
radius of a nucleus $A$ is denoted by $R_A$.
}
\label{projections}
\end{figure}

At very high energies, in a first approximation the produced system
expands only longitudinally \cite{BJORKEN}. If the colliding nuclei
were transversally infinite disks, this would be the case because no
pressure gradients would be generated in the transverse direction.  In
reality, however, the colliding nuclei have rapidly falling nuclear
density distributions (``edges''), and pressure gradients in the
transverse direction are generated in the production and
thermalization stage.  Then, if the lifetime of the collective system
is long enough, a strong transverse expansion develops
\cite{TRANSVERSE}. Consequently, at freeze-out the transverse size of
the system can become clearly larger than the original nuclear radius
$R_A\sim7$ fm for $A\sim200$.  This is illustrated in
Figs.~\ref{projections}. The quantitative details of evolution of the
transverse flow depend on the actual densities and pressure gradients
generated in the very beginning of the QGP phase, and on the actual
EoS in the locally thermal system. The $z,x$-projection in
Fig.~\ref{projections} demonstrates the longitudinal and the
transverse flow.  Notice here that at a fixed time $t$ the different
spatial regions in $z$ are at different stages of the evolution, as
illustrated by Fig.~\ref{spacetime}.

\section{Highlights from the SPS} 

In this section, I have picked up some of the highlights of the
experimental results from URHIC at the SPS. I will concentrate on the
observations that make URHIC very different from hadron-hadron
collisions.  In particular, I will discuss the experimental evidence
of collective behaviour of strongly interacting matter. I will briefly
review the anomalies observed in URHIC at the SPS: the strangeness
enhancement from WA97, the low-mass dielectron enhancement from
CERES/NA45, and the $J/\Psi$-suppression from NA50, and discuss the
physics implications of these experimental facts.

\subsection{Transverse flow}

As discussed in the previous section, if the lifetime of the produced
collective system is long enough, a strong transverse flow is
generated. At freeze-out, the system emits particles from its
decoupling surface according to the appropriate thermal distributions at
each {\em local} freeze-out temperature $T_{\rm fo}$ and baryochemical
potential $\mu_B^{\rm fo}$.  These distributions, however, must be
folded with the (local) flow velocity in each emitting cell. Since the flow is
a collective phenomenon, particles of different masses obtain the same
velocity from it, and consequently the heavier the particle is the
more transverse momentum it gains from the transverse flow.  The
collective transverse flow ${\bf v}_T$ obviously therefore makes the
transverse momentum spectra of heavier final state hadrons flatter
than those of the lighter hadrons. In the $m_T$-distributions
$dN/dm_T^2\sim \exp[-m_T/T_{\rm slope}]$, the inverse slope parameter
$T_{\rm slope}$ can be expressed as
\begin{equation}
T_{\rm slope}=T_{\rm fo}+\frac{1}{2}m\langle v_T^2\rangle
\end{equation} 
at the nonrelativistic domain  $p_T\ll m$ and as a ``blueshift'' formula 
\begin{equation}
T_{\rm slope}=T_{\rm fo}\sqrt{\frac{1+\langle v_T\rangle}{1-\langle v_T \rangle}}
\end{equation} 
at the relativistic region $p_T\gg m$ \cite{BLUESHIFT}.

\begin{figure}[htb]
\vspace{-2cm}
\centerline{\hspace*{-0cm} \epsfxsize=9cm\epsfbox{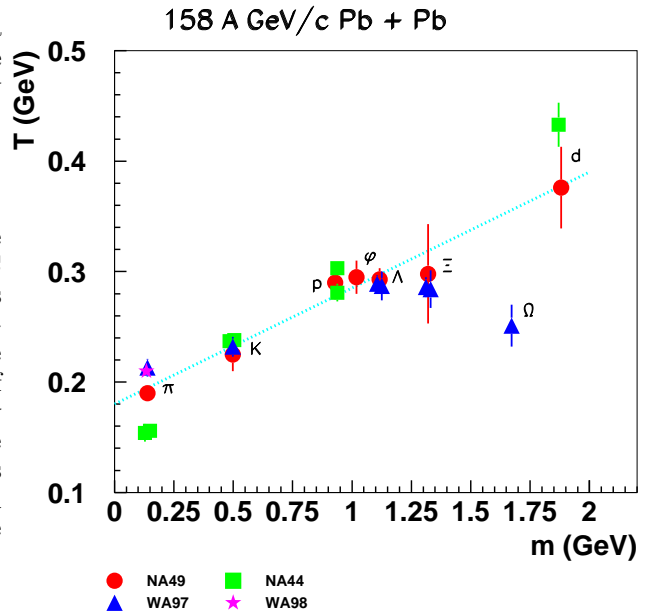}}
\vspace*{-2.5cm} 
\caption{Mass dependence of the inverse slopes $T_{\rm slope}$ of the
$m_T$-distributions. The fit is of the form $dN/dm_T^2\sim
\exp[-m_T/T_{\rm slope}]$ as compiled in \cite{STACHEL}.  The data is
from \cite{ROLANDNA49,NA44T,NA49T,APPELSHAUSER,
WA97T,WA98T} and the figure is from
the SPS-summary \cite{MULLER}.  }
\label{Tslope}
\end{figure}

In Fig.~\ref{Tslope} \cite{STACHEL,MULLER}, also shown by Helstrup (WA97)  
in this conference, the observed $m_T$-spectra of hadrons from NA44
\cite{NA44T}, NA49 \cite{ROLANDNA49,NA49T,APPELSHAUSER}, WA97
\cite{WA97T}, and WA98 \cite{WA98T}, in Pb+Pb collisions at
$E_b=158\,A$GeV, have been fitted to extract the inverse slope
parameters $T_{\rm slope}$ as a function of the particle rest mass.
Provided that the particles freeze out at approximately the same
temperature, $T_{\rm slope}$ should clearly increase with
mass. This indeed seems to be the case, as seen in the figure, with a
notable exception of $\Omega$, perhaps also that of $\Xi$.  It has
been suggested that due to their low interaction cross sections, these 
particles freeze out earlier and thereby benefit less of the transverse
flow \cite{SORGE}. 

\begin{figure}[bht]
\vspace*{-6.cm}
\centerline{\hspace*{0.5cm} \epsfxsize=7.5cm\epsfbox{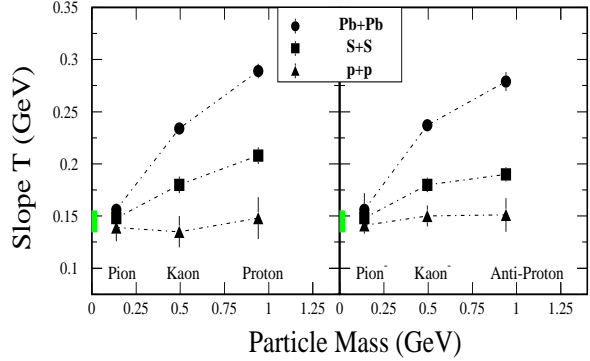}}
\vspace*{0cm}
\caption{The dependence of the inverse slope parameter on the size of
the system as function of mass as observed by the NA44 Collaboration.
The figure is from \cite{NUXU}.}
\label{STsize}
\end{figure}

The average transverse momentum has also been observed to grow with 
the size, as shown in Fig.~\ref{STsize}. Obviously, while the general 
features are consistent with a collective picture,  some details 
i.e. behaviour of $\Omega$ (and $\Xi$?), are not yet fully understood,
and more work remains to be done on the theory.

\subsection{Asymmetric flow}

\begin{figure}[htb]
\vspace{1.5cm}
\centerline{\hspace{-2cm}\epsfxsize=3cm\epsfbox{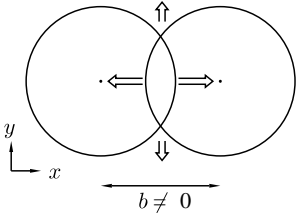}}
\vspace{-1cm}
\caption{A non-central collision projected in the transverse-coordinate plane,
illustrating the initial azimuthal asymmetry in the forming pressure gradients
(the arrows).}
\label{bnonzero}
\end{figure}

A feature in URHIC that in hadron-hadron collisions cannot in practice
be addresses at all, are the non-central collisions, i.e. collisions
with non-zero impact parameter $b$.  For details of the asymmetric
flow effects, I refer to a review \cite{OLLITRAULT}, but to get some
intuition, let us look at Fig.~\ref{bnonzero}, which is a snapshot of
an $A$+$A$ collision projected in the transverse (azimuthal) plane at
the time of maximum overlap of the colliding nuclei. The generated
pressure gradients will obviously be azimuthally asymmetric when
$b\ne0$. Transverse flow is generated by the pressure gradients in
the transverse plane, so it will evidently be azimuthally asymmetric:
$\langle v_x \rangle \ne \langle v_y \rangle$. This in turn causes the
azimuthal angle distributions of final state hadrons to be asymmetric.

\begin{figure}[htb]
\vspace{-0.5cm}
\centerline{
\hspace{-2cm}\epsfxsize=4cm\epsfbox{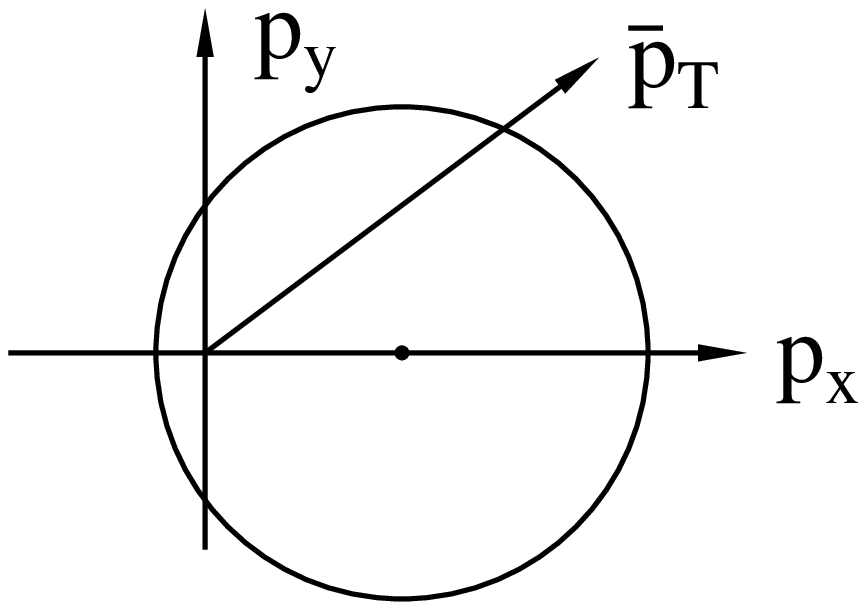}
\hspace{0.5cm}\epsfxsize=4cm\epsfbox{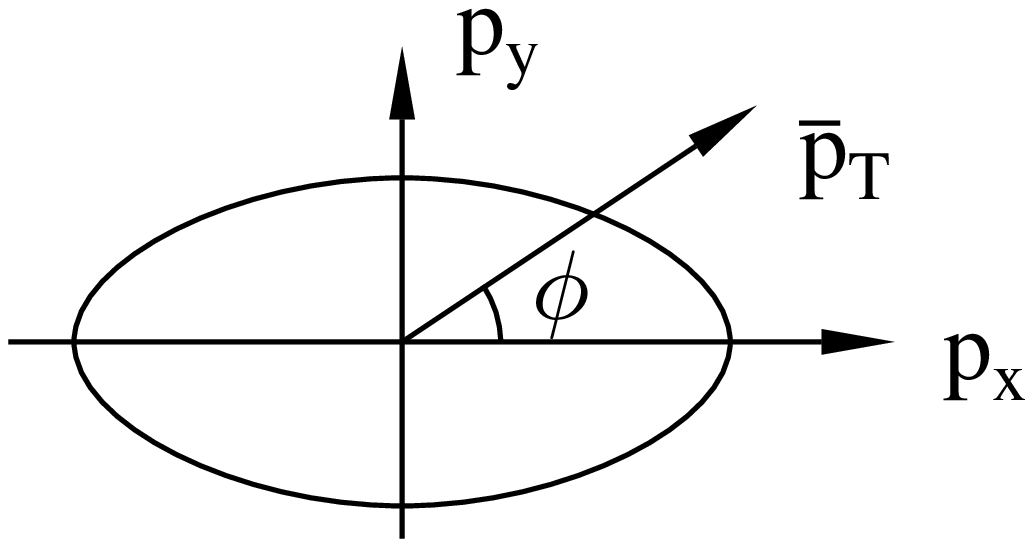} }
\vspace*{-1cm}
\caption{Purely directed (at left) and purely elliptic (at right) flow 
patterns as the azimuthal asymmetries of the transverse momentum
distributions.}
\label{asymmflow}
\end{figure}

To see the azimuthal asymmetry in the measured transverse momentum
distributions, a Fourier analysis is performed, and a fit of the form
\begin{equation}
F(\phi) \sim 1 + 2v_1 \cos\phi + 2v_2\cos2\phi + ...
\end{equation}
can be used in each bin of rapidity. In practice, it suffices to
consider the first two coefficients $v_1$ and
$v_2$. Figs.~\ref{asymmflow} illustrate the two types of asymmetry:
purely directed flow with $v_1\ne0$ and $v_2=0$ and purely elliptic
flow with $v_2\ne0$ and $v_1=0$. The measured angular distributions
are then a mixture of these two effects.

\begin{figure}[bht]
\vspace*{-3.5cm}
\centerline{\hspace*{-2cm} \epsfxsize=7cm\epsfbox{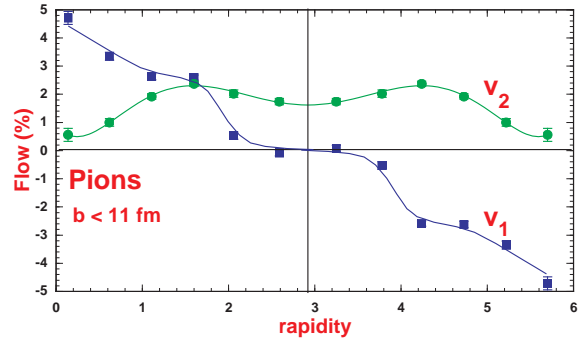}}
\vspace*{-2cm}
\caption{The observed azimuthal asymmetries in the transverse momentum
distributions of pions 
in Pb+Pb collisions at $E_b=158\,A$GeV as function of rapidity and with
impact parameters $b<11$ fm, as measured and analysed by NA49
\cite{POSKANZER}.}
\label{flowna49}
\end{figure}

With the high precision data coming from the SPS, it has indeed now
become possible to observe the few percent asymmetries in the azimuthal
angle distributions. NA49 collaboration has measured the asymmetries
in pion and proton distributions as functions of rapidity
\cite{POSKANZER97,POSKANZER}. As an example, the recent results from
\cite{POSKANZER} for the Fourier coefficients $v_1$ and $v_2$ for the
pion spectrum are shown in Fig.~\ref{flowna49}. For the
results from NA52, see \cite{NA52flow} and from WA98, see
\cite{WA98flow}.

The existence of elliptic flow at the central rapidity (where the
directed flow is zero) is a signature of pressure, and thereby
collectivity, in the system. Once this signature is found, it is
useful to scan the effect as a function of impact parameter, which
controls the energy density reached. An abrupt behaviour as a function
of $b$ would be a direct signature of a QCD phase transition
\cite{HEISELBERG}.  In the (preliminary) analysis \cite{POSKANZER} of
the NA49 data no unexpected behaviour has been observed so far.

\subsection{Particle ratios and thermal models}

Above we saw how the experimental results support the existence of
flow and collectivity in the system.  The flow effects are expected to
change the momentum distributions of final state hadrons but the total
(full coverage $4\pi$) yields of each hadron species should not,
however, be sensitive to the changes in the differential
distributions. In fact, it can be shown that if all the
fluid cells freeze out at the same temperature $T_{\rm fo}$ and at the
same baryochemical potential $\mu_B^{\rm fo}$ everywhere, the
cancellation of flow effects in the particle ratios is complete.
Under these conditions, the underlying dynamics of the system can be
ignored, and purely thermal models can be used to predict the particle
ratios.  The results should then depend only on $T_{\rm fo}$ and
$\mu_B^{\rm fo}$, and by fitting the measured (full coverage) particle
ratios with the predictions of thermal models, $T_{\rm fo}$ and
$\mu_B^{\rm fo}$ can be extracted.

Quite obviously, however, the particle spectra cannot be measured with
a $4\pi$-coverage but some extrapolation is always needed to get the
actual $4\pi$-particle ratios. Secondly, the system is not likely to
freeze out with the same $T_{\rm fo}$ and $\mu_B^{\rm fo}$ everywhere,
(in the forward and backward region the net baryon density is more than in
the middle), so when discussing a thermal model fit to the data, one
effectively considers a substitution of the local quantities $T_{\rm
fo}(x)$ and $\mu_B^{\rm fo}(x)$ by global averages $\langle T_{\rm
fo}\rangle$ and $\langle \mu_B^{\rm fo}\rangle$.  The good news,
however, is that the hydrodynamical simulations \cite{SOLLFRANK} show
that only ${\cal O}(10-15\%)$ deviations from $\langle T_{\rm
fo}\rangle,\langle \mu_B^{\rm fo}\rangle$ are expected at the SPS
energies. This implies that thermal model approach is a reasonable first 
approximation for extracting the freeze-out parameters
$T_{\rm fo}$ and $\mu_B^{\rm fo}$. For more discussion and references, 
see e.g. \cite{HEINZqm99,CLEYMANS}.

\begin{figure}[htb]
\vspace{-5cm}
\centerline{\hspace*{1cm} \epsfxsize=7cm\epsfbox{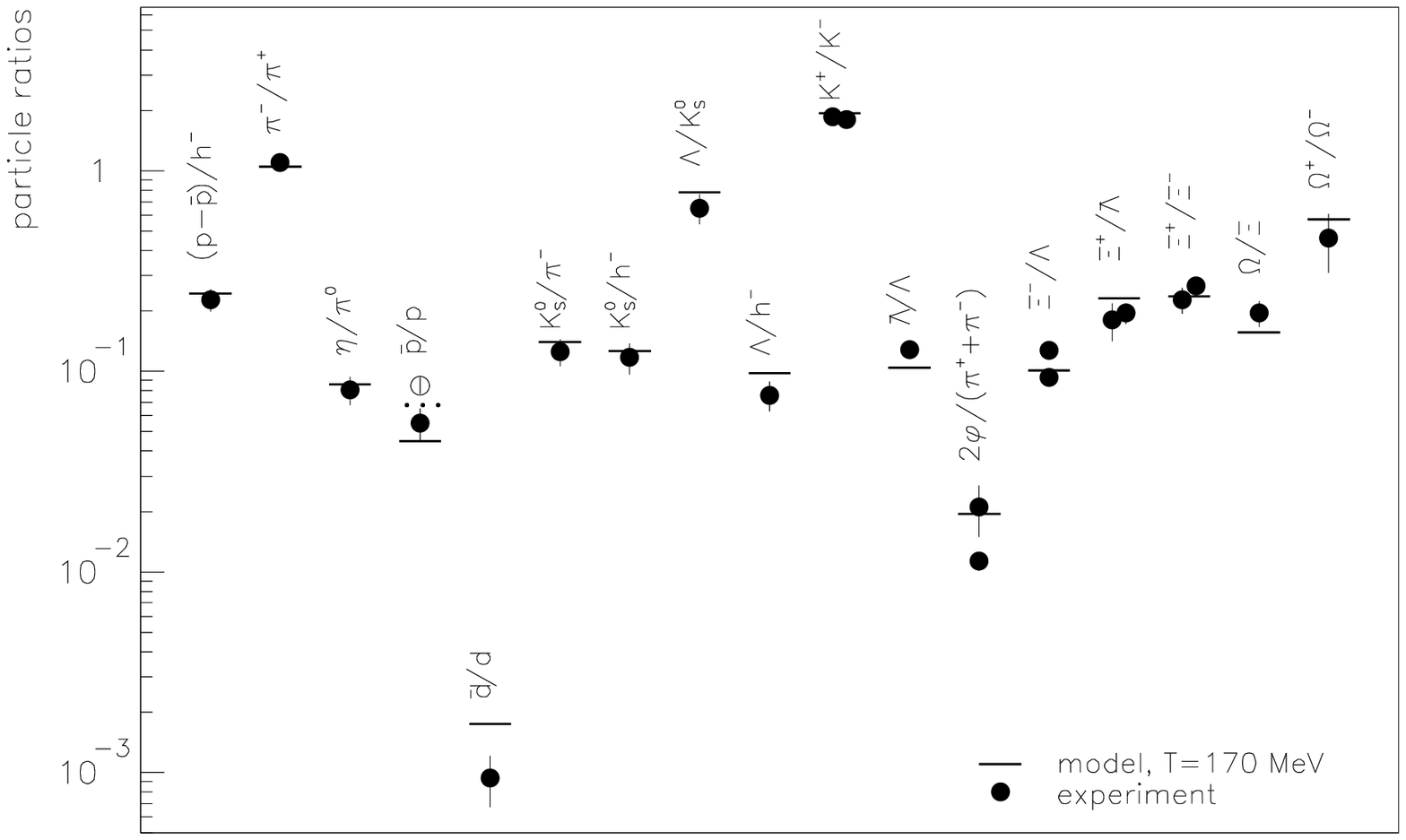}}
\vspace*{0.5cm}
\caption{ Particle ratios as obtained in a thermal model \cite{HEPPE99}
with $T_{\rm fo}=170$ MeV and $\mu_B^{\rm fo}=270$ MeV
as a best fit to the measured particle ratios from 
NA44, NA49, WA97, WA98 and NA50.
}
\label{thermalmodel}
\end{figure}

As an example of a recent thermal model analysis, the results for
particle ratios from \cite{HEPPE99} are shown in
Fig.~\ref{thermalmodel}.  The analysis is straightforward: a
grand-canonical ensemble is used to describe a thermal system of
hadrons at a certain global temperature $T_{\rm fo}$, in a volume $V$ and
at (global) chemical potentials $\mu_B^{\rm fo}$, $\mu_S$,
$\mu_{I_3}$.  Conservation of net baryon number, strangeness and
charge, correspondingly, fixes $V$, $\mu_S$ and $\mu_{I_3}$, leaving
$T_{\rm fo}$ and $\mu_B^{\rm fo}$ as the parameters which are fitted
to an extensive collection of data.  In addition, the hadronic
interactions must be modelled into the partition function, and one should also
allow for decays as well (see \cite{HEPPE99} for details). The best fit
to the data from NA49
\cite{ROLANDNA49,GUNTER,APPELSHAUSER,GABLER,NA49T}, NA44 \cite{NA44T},
WA98 \cite{PEITZMANN}, WA97 \cite{ANDERSEN,KRALIK} and NA50 \cite{JOUAN},
gives $T_{\rm fo}=170$ MeV and $\mu_B^{\rm fo}=270$ MeV.  

The results shown in Fig.~\ref{thermalmodel} indicate that the
measured particle ratios can indeed be quite well described by a
system in thermal and chemical equilibrium.  Thus a picture of a
collective system in thermal and chemical equilibrium is emerging. The
remaining differences between the model and the data are within the
systematic uncertainties of the model and between different data sets
\cite{HEINZqm99}.  I should also mention that there has been a lot of
discussion whether the total strangeness is in chemical equilibrium or
not \cite{RAFELSKI}. In light of the analysis \cite{HEPPE99},
additional fugacities are not necessary.

\subsection {Identical particle interferometry}

The main goal of identical particle interferometry, analogous to
Hanbury Brown-Twiss (HBT) interferometry of stellar objects
\cite{HBT}, is to extract a space-time picture of the system at
freeze-out. By using the measured correlation functions in the
momentum spectra of two (or more) identical pions, it is possible to
estimate the emission volume (homogeneity volume), lifetime of the
system and duration of the emission process (i.e. freeze-out). Also
information on collectivity, i.e. transverse flow, is
obtained. For more detailed reviews of HBT-theory and measurements in
URHIC, I refer to \cite{HEINZqm96,PRATT97,WIEDEMANN99,UAWUH99},  
here I will only describe the basic idea.

\begin{figure}[htb]
\vspace{-4cm}
\centerline{\epsfxsize=7cm\epsfbox{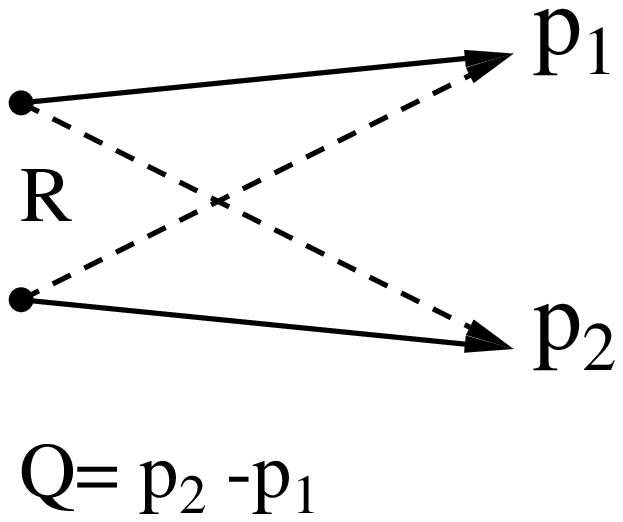}\hspace{5cm}}
\vspace{-5cm} 
\centerline{\hspace{3cm}\epsfxsize=7cm\epsfbox{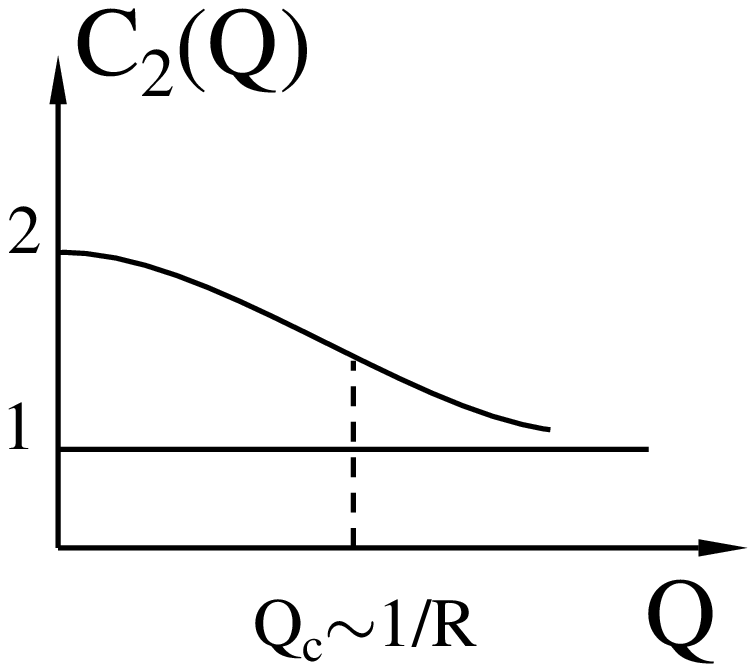} }
\vspace{-2cm}
\caption{Left: Emission of two pions from two sources separated by $R$
(a coordinate 4-vector). Right: The width of the corresponding (measured) 
two-pion 
correlation  function $C_2(Q)$ reflects $R$: $Q_c\sim 1/R$.
}
\label{HBT}
\end{figure}

Let us imagine an emission of two pions with momenta $p_1$ and $p_2$
where the sources are a distance $R$ apart. Relative momentum is
denoted by $Q$, as in Fig.\ref{HBT}, left panel, and the total
momentum by $K$.

The two-particle correlation function is defined as
\begin{equation}
C_2(p_1,p_2) \sim \frac{dN/d^3p_1d^3p_2}{dN/d^3p_1\, dN/d^3p_2}
\end{equation} 
and it can be directly measured. The one-particle distributions (which are
measured) are expressed in terms of space-time integrals of
source functions $S(x,p)$: $dN/d^3p_1= \int d^4xS(x,p)$.  The
source functions contain all the information of the system at
freeze-out, including the transverse flow.  The non-trivial task is to
unfold this information by using the measured correlation functions.

The two-boson correlation function can be written 
in terms of the source functions as \cite{SHURYAK73} 
\begin{equation}
C_2(Q, K) = 1+ \frac{ |\int d^4x e^{iQ\cdot x}S(x,K)|^2 }
{\int d^4xS(x,p_1) \int d^4yS(y,p_2)}.
\end{equation}
The quantitative details naturally depend somewhat on the physical
assumptions made for the functional form of the source function but the
basic idea remains the same: the source radius can be determined 
from the width $Q_c$ of the measured correlation function,   
$Q_c\sim 1/R$, as illustrated in Fig.~\ref{HBT}.

In particular, it should be noted that unlike emission from the stars, 
which are static objects, emission from URHIC is a dynamic process. 
Consequently in HBT for URHIC, the spatial and temporal dimensions are  
non-trivially mixed, and several different radius parameters have to be 
defined. Usually $K$ is chosen to be in the $x,z$ plane with the beam 
in $z$-direction. Then the radius parameters are $R_L$, $R_{out}$ and 
$R_{side}$ in the $z$-, $x$- and $y$-direction, correspondingly. 

\begin{figure}[htb]
\vspace{-3.5cm}
\centerline{\hspace{-1cm}\epsfxsize=10cm\epsfbox{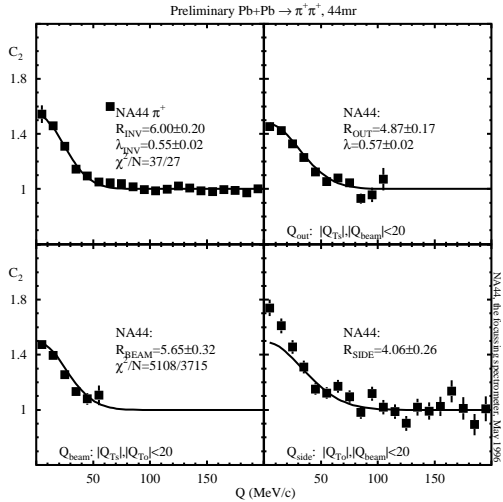}}
\vspace{-4cm}
\caption{An example of the measured identical two-particle correlation 
functions in pion momentum distributions, measured by NA44 \cite{BEARDENHBT}
in Pb+Pb collisions.}
\label{NA44HBT}
\end{figure}

As an  example of an HBT-measurement, let me show the one by NA44
in Fig.~\ref{NA44HBT} \cite{BEARDENHBT}. The remarkable physics
point here is that a typical scale ${\cal O}(20\,{\rm MeV})\sim {\cal
O}(10\,{\rm fm})$ is clearly seen in the high precision data of
two-particle correlation functions. For the most recent experimental
HBT analyses, see the results from NA44 \cite{BEARDEN99}, 
NA49 \cite{GANZ99}, WA98 \cite{VOROS99}, and the talk by Ganz
(NA49) in this conference.

As reviewed in \cite{WIEDEMANN99}, when space-time information from
HBT and complementary information from the $m_T$-distributions are
combined, the freeze-out temperature of the system produced in Pb+Pb
collisions is found to lie in the range $80\,{\rm MeV}<T_{\rm
fo}<120\,{\rm MeV}$, a strong transverse flow is seen to develop,
$0.35<v_T<0.5$, together with a large transverse extension $R_{\rm
fo}\sim 2R_{\rm Pb}$, and lifetime $\sim 10\,{\rm fm}/c$.  In
comparison with the thermal model results in the previous section, a
clear hierarchy is observed: {\it thermal} freeze-out observed in the
HBT and in the transverse momentum distributions takes place later,
i.e. at lower $T$, than {\it chemical} freeze-out which is in turn
reflected by the hadron abundancies.  This is consistent with the
expectation that the chemical reaction rates are much smaller than the
elastic ones.

\subsection{Event-by-Event analysis}

Large multiplicities in single central events allow the study of
fluctuations of different observables on an event-by-event (EbyE)
basis. Also this is a feature of URHIC genuinely different from
hadron-hadron collisions. Naturally, the primary goal of the EbyE
physics is to observe differencies between the events with
and without QGP-formation, as the QCD phase transition may cause large
fluctuations which could show up in measurable quantities like
multiplicities, $\langle p_T\rangle$, particle ratios, etc in single
events. Especially, if the QCD phase transition is of second order,
one might directly observe the critical fluctuations associated with
the phase transition. Also, if some more exotic physics takes place,
such as formation of large enough domains of disoriented chiral
condensates (DCC) \cite{RANDRUP}, the fluctuations from one event 
with DCC to another without could be seen.

\begin{figure}[htb]
\vspace{4cm}
\centerline{\hspace{-0.5cm}\epsfxsize=8.5cm\epsfbox{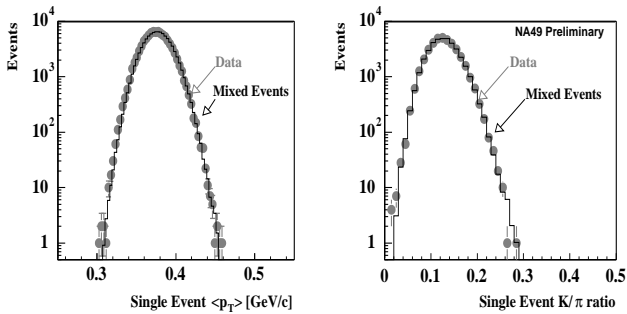}}
\vspace{-3cm}
\caption{Event-by-event fluctuations of single event
$\langle p_T\rangle$ (left) and single event K$/\pi$-ratio (right),
measured by NA49 in Pb+Pb collisions at the SPS.
The dynamical event-by-event fluctuations are $<1\%$ in $\langle p_T\rangle$
and $<4.9\%$ in K$/\pi$. The figure is from \cite{SIKLER}.
}
\label{EbyE}
\end{figure}

Fig.~\ref{EbyE} shows the recent EbyE results from NA49 \cite{SIKLER}.
So far no evidence of non-trivial dynamical fluctuations,
i.e. fluctuations beyond finite statistics and variations of impact
parameter, has been observed in the Pb+Pb collisions at SPS: the
dynamical event-by-event fluctuations are less than 1\% in single
event $\langle p_T\rangle$ and less than 4.9\% in K$/\pi$-ratio (at 90
\% confidence level). It will indeed be very interesting to see what
the corresponding results will be in URHIC at RHIC and LHC/ALICE.

\subsection{Strangeness enhancement}

The observables discussed in the previous subsections support
the interpretation of the the system produced in Pb+Pb collisions at the SPS 
as a collective strongly interacting system. Let me now move on to other 
experimental facts, namely the observed anomalies. First, let us have a 
look at the strangeness enhancements observed. For more detailed reviews 
and for more complete list of references to the measurements of 
strangeness production in URHIC, see e.g. \cite{ODYNIEK,SQM98}.

In comparison to hadron-hadron and e$^+$+e$^-$ collisions, a {\em global
enhancement} of strangeness production has been observed in URHIC.
At the SPS, the first observation of this was made by NA35 \cite{NA35}
in S+S collisions at $E_b=200\,A{\rm GeV}$.
The global enhancement can be quantified by defining a
parameter which counts the strange quark-antiquark pairs produced 
relative to the $u\bar u$ $d\bar d$ produced \cite{BECATTINI}
\begin{equation}
\lambda_s^{AA} \equiv \frac{2\langle s+\bar s\rangle}{\langle u+\bar
u\rangle + \langle d+\bar d\rangle}\approx 2\lambda_s^{pp}
\end{equation}

\begin{figure}[htb]
\vspace{-1cm}
\centerline{\hspace{0cm}\epsfxsize=7cm\epsfbox{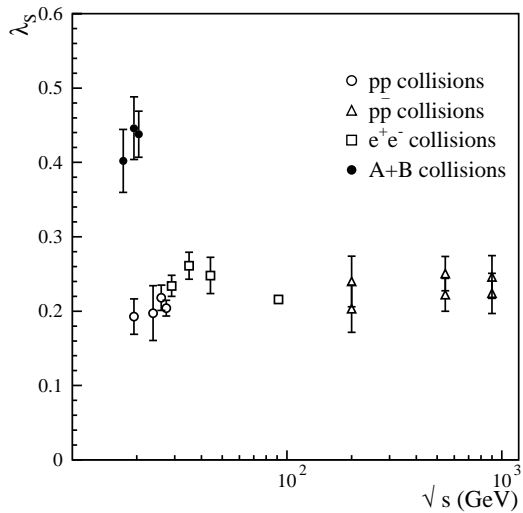}}
\vspace{-1.5cm}
\caption{The global strangeness is observed to be enhanced by a factor
$\sim 2$ in $A$+$B$ (S+S, S+Ag and Pb+Pb)
relative to e$^+$+e$^-$,  p+p and p+$\bar {\rm p}$
collisions. The figure is from \cite{BECATTINI}.  
}
\label{globalS}
\end{figure}

This indicator is plotted in Fig.~\ref{globalS} 
for S+S and S+Ag at $E_b=200\,A{\rm GeV}$
based on the data from NA35 \cite{NA35more},
and for Pb+Pb collisions at $E_b=158\,A{\rm GeV}$
based on the measurements by NA49 
\cite{JONESNA49,NA49s} and compared with the corresponding
compilation of data from e$^+$+e$^-$, p+p and p+$\bar{\rm p}$ collisions
(for the refs, see \cite{BECATTINI}). The observation is that
$\lambda_s^{AA}\approx2\lambda_s^{pp,ee}$: the global strangeness
enhancement is thus a unique feature of URHIC.

\begin{figure}[htb]
\vspace{-2cm}
\centerline{\hspace{1cm}\epsfxsize=7.5cm\epsfbox{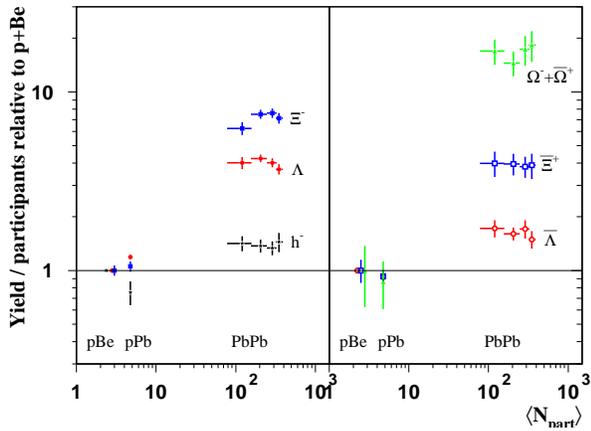}}
\vspace{-2.5cm}
\caption{The specific strangeness enhancement as measured by WA97.
The yields per participant per unit rapidity relative to p+Be
as a function of the number of participants (centrality)
\cite{ANTINORI}.}
\label{specificS}
\end{figure}

In addition to the global enhancement, also {\em specific
enhancements} in the production of $K,\bar K, \Lambda,\bar \Lambda,
\Xi, \bar\Xi, \Omega, \bar \Omega$ and $\phi$ at midrapidity in Pb+Pb
collisions relative to that in p+Be collisions have been reported by the WA97
\cite{ANTINORI, ELIA}, NA49 \cite{SIKLER,HOHNE} and NA50 \cite{NA50s}
collaborations.  The multistrange hadron yields, in particular, are
strongly enhanced in central Pb+Pb collisions, as seen in
Fig.~\ref{specificS}, where the yields per participants relative to
pBe, measured by WA97, are plotted as a function of the average number
of participants in p+Pb and Pb+Pb collisions at the SPS
\cite{ANTINORI}. For more details, see the talk by Helstrup (WA97) in
this conference.

Notice especially that the number of negative hadrons scales with the
number of participants, and relative to the charged hadrons the
enhancement in $\Omega^-+\Omega^+$ yield is roughly a constant factor
10. With the 1998 run, the Pb+Pb event statistics was practically
doubled, and WA97 will soon cover the region below 100
participants (see Helstrup and \cite{ANTINORI}). It will certainly be
very exciting to see whether there is a threshold in the number of
participants (centrality) after which the strong specific strangeness
enhancement takes place.

What are the implications of these experimentally observed facts? If
the produced dense system consists of {\em uncorrelated} quarks (and
antiquarks), it is intuitively clear that the more a hadron contains
strange quarks (or antiquarks) the more it benefits of the global
strangeness enhancement in hadronization.  This picture is quantified
in statistical hadronization (quark coalescence) models \cite{BIALAS},
which seem to work for $A$+$A$ collisions but not for p+$A$
collisions.  On the other hand, hadronic kinetic models do not
reproduce the data on multistrange hadrons due to high mass-thresholds
and low chemical reaction cross sections. Also approaches with string
breaking followed by hadron rescattering models (see \cite{ANTINORI}
for the actual comparison and for further references) cannot reproduce
the centrality pattern observed. The implication of the observed
strangeness enhancement therefore is that the quark degrees of freedom
are indeed essential, and that strangeness must have been produced
very early on in the collision, in the pre-hadronic stage.

From the thermal models (Fig.~\ref{thermalmodel}), we remember that
the particle ratios, including those with multi-strange hyperons are
close to chemical equilibrium values.  This, together with the
conclusion of the importance of the quark degrees of freedom, 
points towards thermalized QGP.

\subsection{Low-mass $e^+e^-$ enhancement}  

Another very interesting anomaly observed in URHIC at the SPS is the
excess of electron-positron pairs in the mass region $250 \,{\rm MeV}
\lsim M_{e^+e^-}\lsim 700 \,{\rm MeV}$, usually referred to as the
``low-mass'' region. The excess in $e^+e^-$ production was first observed by
NA45/CERES in S+Au collisions at $E_b=200\,A{\rm GeV}$
\cite{NA45,CERES95,TSERRYA95}, and a similar excess was observed in
$\mu^+\mu^-$ by the HELIOS-3 collaboration in S+W collisions at
$E_b=200\,A{\rm GeV}$ \cite{HELIOS3}, and by NA38 \cite{NA38} in S+U
at $E_b=200\,A{\rm GeV}$. The low-mass excess has been confirmed by
CERES for heavier systems in Pb+Au collisions at 
$E_b=158\,A{\rm GeV}$ \cite{TSERRYA97,LENKEIT}.

\begin{figure}[htb]
\vspace{-0cm}
\centerline{\hspace{0.5cm} \epsfxsize=7.cm\epsfbox{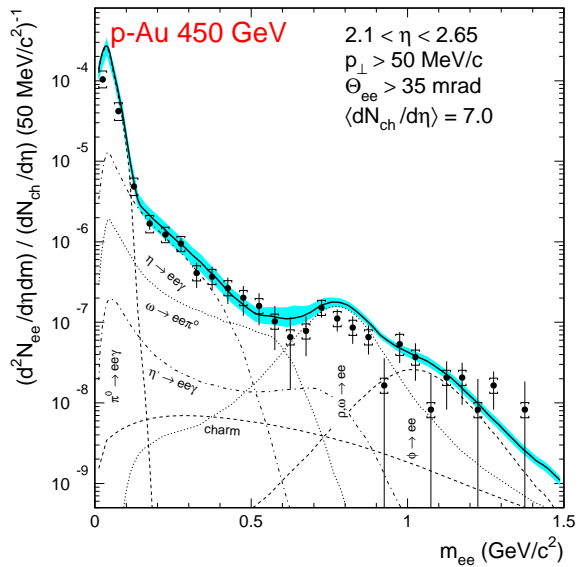}}
\vspace{-1.5cm}
\caption{Inclusive mass distribution of $e^+e^-$-pairs in p+Au 
collisions at 450 GeV scaled with charged particle multiplicity 
\cite{CERESpA}. The curves are the known sources from hadron decays.
}
\label{lowmasspAu}
\end{figure}

The reference data is given by the $e^+e^-$ production in p+Be and
p+Au collisions at $E_p=450$ GeV: no enhancement is observed there,
the data can be explained by including contributions from all the known
hadron decays.  The mass distributions of $e^+e^-$-pairs measured in p+Au by
CERES \cite{CERESpA} and scaled by the measured charged particle
multiplicity (within the CERES acceptance) are shown in
Fig.~\ref{lowmasspAu}. The cocktail of the several different sources
of $e^+e^-$-pairs is shown, and it is seen to reproduce the measured
data within the estimated errors (the shaded band) quite well.

\begin{figure}[htb]
\vspace{-1cm}
\centerline{\hspace{0cm}\epsfxsize=8cm\epsfbox{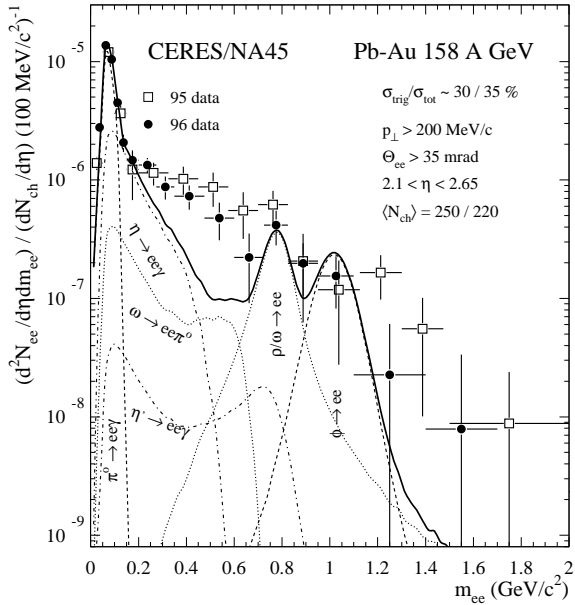}}
\vspace{-0cm}
\caption{Low-mass enhancement of $e^+e^-$-pairs in Pb+Au collisions 
at $E_b=158\,A$GeV \cite{LENKEIT}.}
\label{lowmassPbAu}
\end{figure}

However, when one extrapolates the cocktail-plot to S+Au and Pb+Au
collisions, a clear enhancement is observed, as shown in
Fig.~\ref{lowmassPbAu} \cite{LENKEIT} for the mass distributions of
$e^+e^-$-pairs scaled by the charged particle multiplicity.  In the
figure, data sets from '95 and '96 are compared with the sum of the
expected contributions from hadron decays. In the cocktail-plot the
particle ratios are taken from a thermal model fitted to measured
ratios in Pb+Pb ($T_{\rm fo}=$ 175 MeV and $\mu_B=$ 270 MeV,
fig.\ref{thermalmodel}), and the $\eta$-and $p_T$-distributions follow the
measured systematics in Pb+Pb collisions.  The enhancement in the
mass-region $250 \,{\rm MeV} \lsim M_{e^+e^-}\lsim 700 \,{\rm MeV}$ is
observed to be $2.6\pm0.5 ({\rm stat})\pm0.6({\rm syst})$ for the '96
data set, and $3.9\pm0.9 ({\rm stat})\pm0.9({\rm syst})$ for the '95
data set \cite{LENKEIT}. The excess has been observed to concentrate
at low pair transverse momentum.

Interestingly, the excess scales more strongly than linearly with
charged particle multiplicity, suggesting that it is due to a
medium-effect.  The shape of the mass-spectrum is not consistent with
$\pi^+\pi^-$ annihilation in free space, either, which again supports
a medium-effect as the origin of the low-mass enhancement. There has
been considerable theoretical activity in trying to explain the
effect. For recent reviews, see \cite{WAMBACH97,RAPP99,TSERRYA97}.
The suggestions vary from a collisional broadening of the $\rho$-meson
(i.e. $\rho$ has a shorter lifetime in a dense hadronic medium, which
leads to a broader peak in the mass distribution) and in-medium
modifications of the $\rho$ spectral function \cite{WAMBACH97,RAPP99}
to a possible change in the mass of $\rho$ \cite{BROWN}. 

\begin{figure}[htb]

\centerline{\hspace{0cm}\epsfxsize=6cm\epsfbox{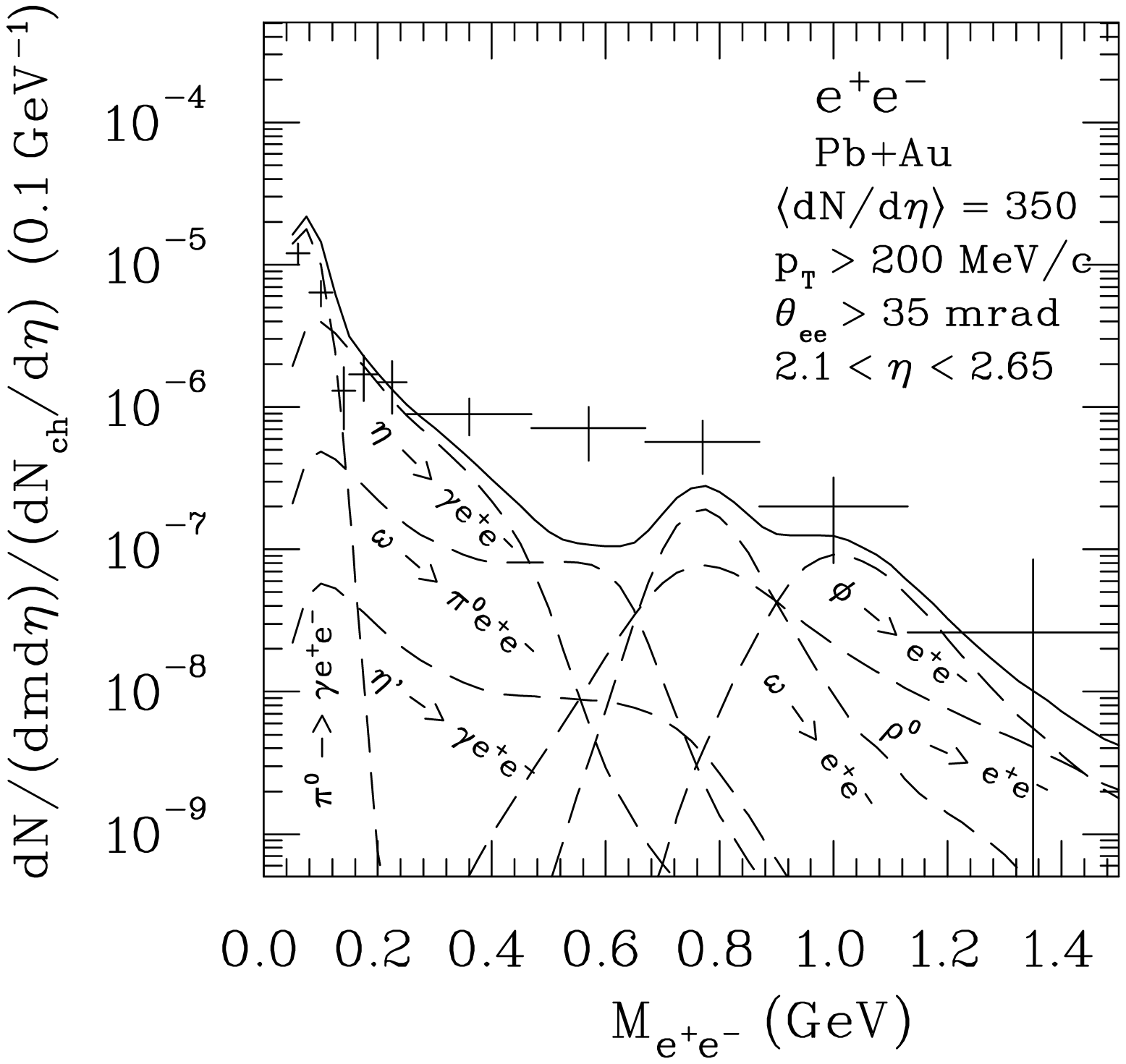}}
\vspace{1cm}
\centerline{\hspace{0cm}\epsfxsize=6cm\epsfbox{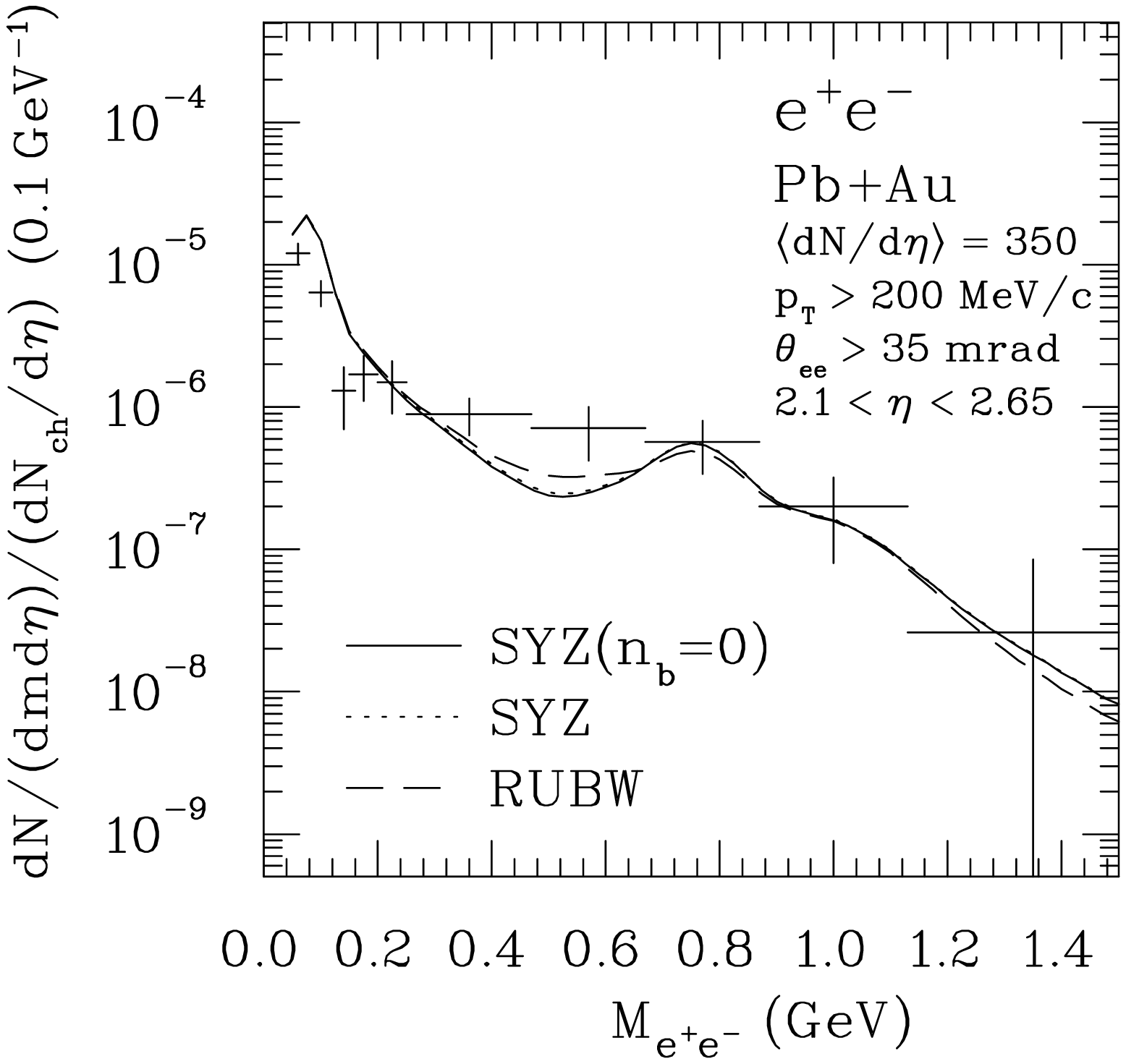}}
\caption{Mass spectrum of $e^+e^-$-pairs scaled with charged particle
multiplicity as predicted by a hydrodynamic simulation \cite{HUOVINEN}. 
Upper panel: contribution from the hadron decays. Lower panel: 
total yield, which includes the thermal emission.
}
\label{hydroresult}
\end{figure}

Let me also show the dielectron results from \cite{HUOVINEN} based on
a hydrodynamical simulation where the whole space-time evolution of
the system is taken into account \cite{HYDRO}. In this approach, the
total yield of dielectrons (and dileptons in general) consists of the
pairs from the decays of resonances originating from the freeze-out
surface (see Fig.~\ref{spacetime}), and of the thermal dielectrons
emitted from the dynamically evolving fireball throughout its whole
space-time evolution. The measured hadronic momentum spectra are first
reproduced to constrain the hot initial state and the EoS
simultaneously \cite{HYDRO}. After this, the emission of thermal
$e^+e^-$-pairs can be predicted by using the thermal rates
\cite{THERMALRATES,RAPP99} at each local $T$ and $\mu_B$.  The
$e^+e^-$ contribution from the resonance decays alone is shown in
Fig.~\ref{hydroresult} by the upper panel. Adding the thermal pairs
gives the lower panel. We observe that that thermal emission
dominates the total yield of $e^+e^-$-pairs in the region of the
experimentally observed excess. Let me also remind that the low-mass
enhancement is {\em not} considered to be a direct signal of QGP:
contribution of thermal emission from the QGP-phase is negligible in
the region of the low-mass enhancement (see \cite{HUOVINEN}).

Although theoretically still not fully resolved, it is evident that
with the low-mass enhancement of $e^+e^-$-pairs measured by CERES, we
are having a first glimpse into the dynamics of mesons in baryon-rich
environment exceeding a temperature $T\sim120$ MeV.  In the near
future, CERES will collect more data, also with the lower energy beams
($E_b=40\,A$GeV) and include an additional TPC in the experimental
setup \cite{LENKEIT}. With improved statistics, and a better
mass-resolution and signal-to-background ratio, CERES will
certainly shed more light to the details of low-mass dielectron excess 
as a signature of a hot hadronic gas.

Regarding the observed dilepton mass spectrum in general, let
me note that also the intermediate mass dimuons, $1.5\,{\rm GeV} <
M_{\mu\bar\mu} < 2.5$ GeV, show an excess relative to the
conventionally known sources, as observed by NA38 in S+U \cite{RAMOS},
and HELIOS-3 in S+W collisions at $E_b=200\,A$GeV \cite{HELIOS3}, and
NA50 in Pb+Pb at $E_b=158\,A$GeV \cite{BORDALO99}. It will be
interesting to see whether this excess could also be explained by
production of thermal dileptons \cite{RUUSKANENqm97}.

\subsection{$J/\Psi$ suppression}

Of the anomalies observed in URHIC at the SPS, the one that has
generated most excitement, is the anomalous suppression of $J/\Psi$ in
central Pb+Pb collisions at $E_b=158\,A$GeV, observed by the NA50
collaboration \cite{GONIN96,NA5097,RAMELLO97,NA5099,CICALO}. The idea of
studying the suppression of $J/\Psi$ production as a signal of the QGP
originates from Satz and Matsui in 1984 \cite{MS84}: formation of
$J/\Psi$ bound state should be very efficiently Debye screened in
QGP. For details of the NA50 measurements and analysis, see
\cite{CICALO} and the talk by De Falco (NA50) in this conference. For a more
detailed overview of the theory of $J/\Psi$ suppression, see
e.g. \cite{KHARZEEVqm97,VOGT99,SATZqm99} and the talk by Nardi in this
conference.

\begin{figure}[htb]
\vspace{-1.5cm}
\centerline{\hspace{-0cm}\epsfxsize=6cm\epsfbox{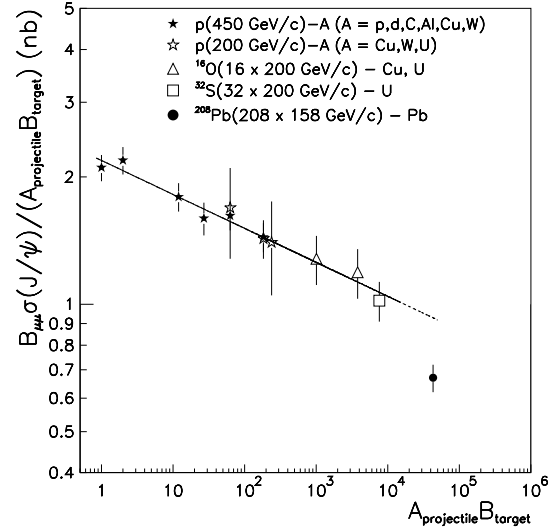}}
\vspace{1cm}
\caption{Suppression of $J/\Psi$ production as a function of 
$AB$: normal nuclear suppression from p+p to p+$A$ and to S+U,
and anomalous suppression in Pb+Pb. The figure is from NA50, ref. 
\cite{NA5097}.}
\label{JPsivsAB}
\end{figure}

The key question naturally is relative to what the $J/\Psi$
suppression appears in the heaviest URHIC systems.  As shown in
Fig. ~\ref{JPsivsAB} \cite{NA5097}, a clear and smooth suppression
pattern is observed from p+p and p+d collisions to p+$A$ and S+U
collisions, as measured by the NA38 and NA51 collaborations
\cite{NUCSUP}: $B_{\mu\mu}\sigma({J/\Psi})$ is decreasing as a function
of $A*B$. For the comparison, the measurements have been rescaled to
the same $\sqrt s$. A scaling $B_{\mu\mu}\sigma({J/\Psi}) \sim
(AB)^{0.92}$ is found \cite{NA5097}. As seen in the figure, the
$J/\Psi$ in Pb+Pb collisions clearly deviate from the pattern of
normal nuclear suppression. The additional suppression of $J/\Psi$ in
Pb+Pb collisions is often referred to as the ``anomalous'' one.

At the same time, the Drell-Yan (DY) dilepton production at is {\em
not} observed to be suppressed \cite{NA5097}\footnote{The Drell-Yan 
computation here assumes that any nuclear effects in the parton
distributions \cite{EKS98} are negligible within the errorbars.}:
$\sigma(DY)_{\rm exp}/\sigma(DY)_{\rm th}\sim AB$.  Obviously then the
suppression of $J/\Psi$ comes from the final state interactions,
i.e. from absorption of the produced bound state within normal nuclear
matter (the beam and target). The nuclear absorption, often called as
``normal'' suppression, can be understood in Glauber-type models, see
\cite{GERSCHEL,KHARZEEV97} for details. For discussion of the
absorption by hadronic comovers, see \cite{GAVIN}.  The observed
nuclear absorption pattern (Fig.~\ref{JPsivsAB}) can be reproduced by
using absorption cross-sections of the $J/\Psi$ (or its pre-hadronic
state) of the order of 6...7 mb \cite{NA5097,KHARZEEV97}

\begin{figure}[htb]
\vspace{2.cm}
\centerline{\hspace{-2cm}\epsfxsize=10cm\epsfbox{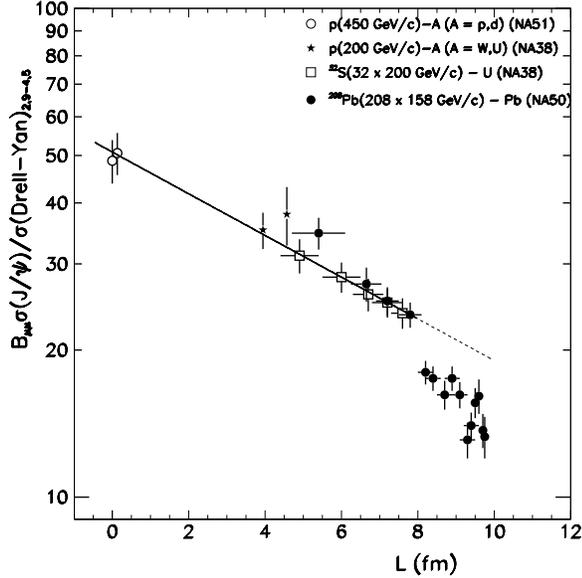}}
\vspace{-3.5cm}
\caption{The suppression of $J/\Psi$ over Drell-Yan pairs in
$2.9\,{\rm GeV}< M <4.5$ GeV as a function of the average nuclear path length 
$L$ of the  $c\bar c$ pre-resonance \cite{NA5099}, for NA38, NA50 and 
NA51 data. The NA50 Pb+Pb data (filled circles) is from the 1996 run.}
\label{PSIDYvsL1}
\end{figure}
\begin{figure}[htb]
\vspace{2.5cm}
\centerline{\hspace{-2cm}\epsfxsize=13cm\epsfbox{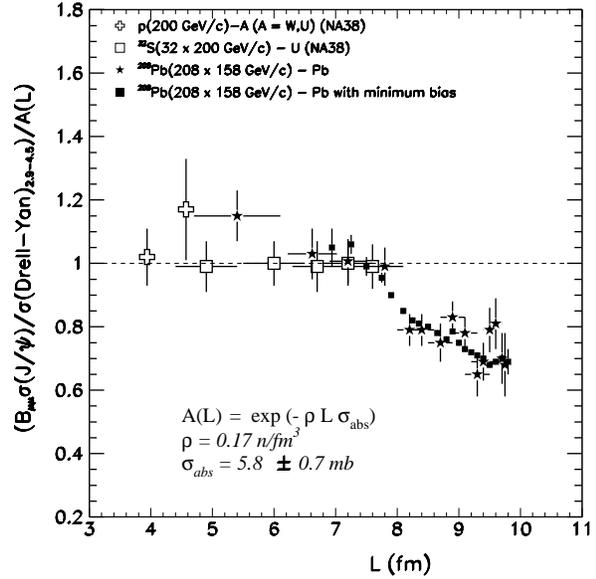}}
\vspace{-7.cm}
\caption{The same as in the previous figure but the with the normal
nuclear suppression divided out \cite{NA5099}. The Pb+Pb data is from 
1996 run, and also the new minimum bias analysis is shown (the 
filled squares). In comparison with the previous figure, notice here 
the different scale in $L$.}
\label{PSIDYvsL2}
\end{figure}
\begin{figure}[htb]
\vspace{0cm}
\centerline{\hspace{-0.5cm}\epsfxsize=8cm\epsfbox{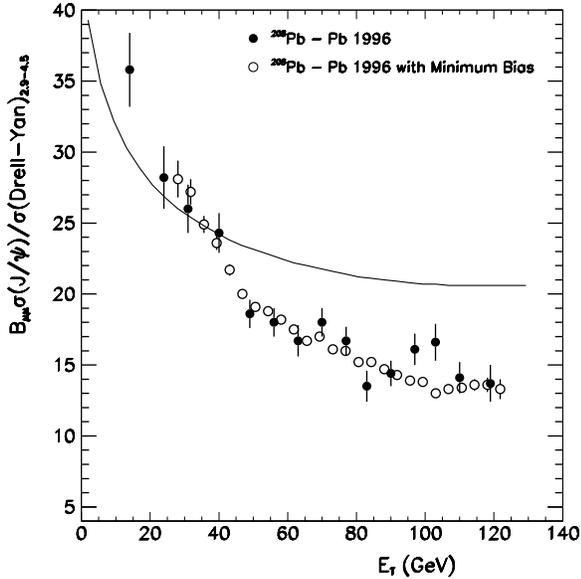}}
\vspace{0cm}
\caption{The ratio $\sigma_{J/\Psi}/\sigma_{\rm DY}$ at $2.9\,{\rm GeV}
<M< 4.5 \,{\rm GeV}$ in Pb+Pb collisions as a function of
$E_T$  \cite{NA5099}.  The 1996 data is analysed in two different ways, 
see  \cite{NA5099} for details.}
\label{PSIDYvsET1}
\end{figure}

NA50 has also carefully studied the suppression as function of
transverse energy $E_T$.  In each Pb+Pb collision, the produced
transverse energy and the energy observed in the zero-degree
calorimeter correspond to an impact parameter $b$. The produced $c\bar
c$-pair bound state which is to form a $J/\Psi$ first has to go
through the rest of the projectile and the target, so for each $b$
($E_T$) there is an average nuclear path length $L$ which the
pre-resonance has to survive. The measured Drell-Yan pairs serve as a
non-suppressed background in each $E_T$-bin, i.e. at each $b$, and
$B_{\mu\mu}\sigma({J/\Psi})/\sigma(DY)$ can be plotted both as a
function of $E_T$ and as a function of $L$, as shown in
Figs.~\ref{PSIDYvsL1}, \ref{PSIDYvsL2} and \ref{PSIDYvsET1}
\cite{NA5099}. The DY cross section $\sigma(DY)$, against which
$B_{\mu\mu}\sigma({J/\Psi})$ is compared, is computed
at $2.9\,{\rm GeV}< M<4.5\,{\rm GeV}$ by using the
measured DY-pairs at $M>4.5$ GeV to get the correct normalization (Note
that equally well one could directly use the measured Drell-Yan pairs;
this has also been done by NA50 but is not shown here).

In Fig.~\ref{PSIDYvsL1}, the filled circles at largest values of $L$
represent the most central Pb+Pb collisions, and the ones at smallest $L$
correspond to impact parameters $\langle b\rangle=10.8$ fm
\cite{NA5099}.  The peripheral Pb+Pb are seen to coincide with 
lighter systems, as one would expect to. It is also
interesting to divide out the normal nuclear suppression and replot
the suppression, as done in \cite{NA5099}, from where I have
borrowed also Fig.~\ref{PSIDYvsL2}.  Notice here the importance of the
p+p and p+d points ({\em not} shown in this figure) in determining the
amount of normal nuclear suppression of $J/\Psi$ at large values of
$L$. The filled circles show the results from the new minimum bias
analysis of NA50 \cite{NA5099}, as discussed by De Falco at this 
conference. The abrupt drop seen in the figure for the $J/\Psi$ suppression
in Pb+Pb is the best candidate so far for a threshold behaviour in URHIC at 
the SPS. Naturally, the hope is that this would be the first direct
observation of the QGP formation in the central Pb+Pb collisions.

Fig.~\ref{PSIDYvsET1} from \cite{NA5099} shows the suppression of
$J/\Psi$ over Drell-Yan directly as a function of the measured $E_T$,
as obtained in the two different analyses of the data from 1996 run.
The 15 $E_T$ bins correspond to the 15 bins in $L$ in the previous
figures and the solid line represents the expectation for the normal
nuclear suppression as shown in the previous figures. Here, one should
again keep in mind the importance of the lighter systems in
determining the solid line. The suppression pattern is, naturally, the
same as in the previous figures.

\begin{figure}[htb]
\vspace{0cm}
\centerline{\hspace{0.2cm}\epsfxsize=7cm\epsfbox{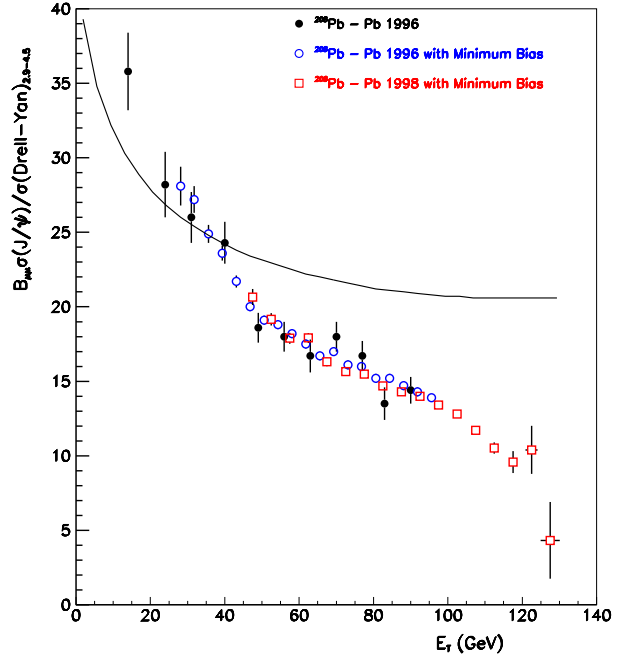}}
\vspace{-1cm}
\caption{ Preliminary analysis of the 1998 data for suppression of
$J/\Psi$ and comparison with the analyses of the 1996 data.
The figure is from \cite{CICALO}. }
\label{PSIDYvsET2}
\end{figure}

Interestingly, as recently reported by NA50 \cite{CICALO}, there may
actually be {\em another drop} in the $J/\Psi$ suppression pattern at the
largest values of $E_T$. The latest (preliminary) analysis of the 1998
data is shown in Fig.~\ref{PSIDYvsET2} from ref. \cite{CICALO}.  As
discussed in \cite{CICALO}, the second drop was not observed in the
data from 1996 run, where a thicker target was used, because of a
bias from re-interaction events. This problem is removed in the 1998 
set up. The possibility for a second drop is indeed very exciting, 
and it would support the prediction in \cite{2DROP} of melting of 
different $c\bar c$ bound states at different energy densities.

For a future improvement, the possibility to add a new vertex detector
into the NA50 setup is discussed.  This suggestion, called
NA6$i$, was also presented in this meeting by Shahoyan. The benefit of
this would be that open charm contribution could be measured more
accurately, and this would serve as a very useful background both for the
$J/\Psi$ suppression measurements and for the intermediate mass muon pairs.

For a compact review of the theoretical interpretations of the
observed $J/\Psi$ suppression, let me suggest \cite{KHARZEEVqm97}, see
also the talk by Nardi in this conference. As one is discussing
formation and interactions of a QCD bound state, perturbative
calculations are hard to do. Consequently, different models have been
presented, see e.g.  \cite{GERSCHEL, GAVIN, KHARZEEV96, KHARZEEV97,
WONG96, BLAIZOT96, PAJARES, VOGT99,SATZqm99}. Several questions, like
the energy dependence of the dissociation cross section of the $c\bar
c$-pair \cite{DSSZ,HOYER}, and applicability of QCD factorization in the
bulk production of $J/\Psi$ in $AA$-collisions have been and are still
discussed.  Even if all the details are theoretically not yet fully known,
there seems to be a general consensus on that while the normal nuclear
suppression of $J/\Psi$ can be accounted for by a more conventional
hadronic picture, to reproduce the anomalous suppression, especially
with the observed abrupt behaviour - possibly even with two drops - an
additional ingredient very efficient in destroying the $c\bar c$ bound
state is needed.  Putting this together with the implications of a
collective system and of the importance of partonic degrees of
freedom, as discussed in the previous sections, it is very likely that
we are now finally looking at a deconfined phase of QCD-matter.

\section{The future: RHIC \& LHC/ALICE}

\subsection{RHIC: Relativistic Heavy Ion Collider}

Until now the fixed target experiments at the SPS at $\sqrt s \sim 20
\,A$GeV have been the URHIC of highest collision energy.  The next
step will be taken by the Relativistic Heavy Ion Collider (RHIC) at
Brookhaven National Laboratory, where $^{197}$Au-beams will be
collided onto $^{197}$Au-beams at $\sqrt s=200 \,A$GeV, and luminosity
$L\sim 10^{26..27}\,{\rm cm}^{-2}{s}^{-1}$.  RHIC will also
collide proton-beams with proton-beams at $\sqrt s= 500$ GeV and
$L=10^{31}\,{\rm cm}^{-2}{s}^{-1}$, and make p+ $A$ and $A$+$B$ collisions
with various $A$ and $B$.  Collisions with polarized protons will also
be done.  At the time of the this meeting, beam tests were being
done in the two rings, with the aim to get the collider in operation
in December 1999\footnote{Some delays can now be foreseen.}
\cite{COMMISSIONING}.

Without going to any details, let me here just list the heavy-ion
experiments at RHIC, together with their main physics goals. For a more
complete review, see e.g. \cite{HARRIS}.  The main goal of RHIC will
be to study the strongly interacting matter produced at URHIC, and to
observe the QGP through various independent signals. Thanks to the
varying beams and cms-energies, RHIC will provide us with an extensive
$\sqrt s$- and $AB$-systematics, which will be important in comparing
the events with and without QGP formation.

There are four major heavy-ion experiments at RHIC, involving {\cal
O}(900) scientists and {\cal O}(80) institutions. There are two
``large'' experiments, STAR (Solenoidal Tracker at RHIC) \cite{STAR}
and PHENIX (Pioneering High Energy Nuclear Interaction Experiment)
\cite{PHENIX}, and two ``small'' ones, BRAHMS (Broad Range Hadron
Magnetic Spectrometer) \cite{BRAHMS} and PHOBOS \cite{PHOBOS}.

The main elements of the STAR experiment are silicon vertex tracker
system (SVT), time projection chamber (TPC), forward radial-drift TPC
(FTPC), time of flight measurement system (TOF) and electromagnetic
calorimeter (EMC). The primary physics goal of STAR is to study
strongly interacting matter at high densities in central $A$+$B$
collisions. This will be achieved by measuring hadrons over a large solid
angle, and measuring the global observables on an event-by-event basis.
STAR will also study peripheral $A$+$B$ collisions and polarized p+p
collisions.

The main building blocks of the PHENIX experiment are multiplicity
vertex detector (MVD), ring-image Cherenkov counter (RICH),
time-expansion chamber (TEC), TOF and EMC. PHENIX will mainly
concentrate on measuring leptons, photons and hadrons to search for
the QGP signatures.

BRAHMS is a forward and mid-rapidity hadron spectrometer.  The
physics goal is to make a systematic study of particle production in
$AA$ collisions from the peripheral to the most central impact
parameters.  BRAHMS should basically answer the question of where the
baryons go.  PHOBOS is a compact multiparticle spectrometer which will
measure single particle spectra and correlations between particles
with low transverse momenta and to characterize events using a
multiplicity detector.

\subsection{ALICE: A Large Ion Collider Experiment}

The ultimate step in the cms-energy of URHIC will be taken by A Large
Ion Collider Experiment, ALICE \cite{ALICE} at the Large Hadron
Collider (LHC) at CERN in 2005. ALICE will be the only detector at the
LHC dedicated for URHIC.

The heaviest beams colliding at ALICE will be $^{208}$Pb+$^{208}$Pb 
at $\sqrt s=5.5\, A{\rm TeV}$. The luminosity in these collisions,  
$L=2\cdot10^{27}\,{\rm cm}^{-2}{s}^{-1}$, corresponds to some 
$10^4$ interactions$/$s. ALICE will also perform  p+p and p+$A$ 
collisions to collect reference data for Pb+Pb, and it will also 
make $A$+$B$ and $A$+$A$ collisions with lighter nuclei to vary the 
produced densities of the QGP. According to the present running 
scenario, the running time for heavy ions will be about 10\% of 
the LHC year.

In comparison with SPS and RHIC, there are definite physics advantages 
in increasing the cms-energy as much as possible:
\begin{itemize}
\item net baryon density near $y\sim0$ is smaller; $\mu_B\ll T$, and the
EoS approaches that of the theoretically better understood $\mu_B=0$ case
\item initial QGP energy density is higher
\item lifetime of the QGP phase and of the whole system is longer
\item freeze-out volume is larger (in a comoving sense, the system never 
decouples simultaneously everywhere)
\item one moves into the applicability region of perturbative QCD (pQCD): 
First, there will in general be higher rates for harder
QCD-probes of the QGP, such as high-$p_T$ jets, direct photons at high
$p_T$, large mass dileptons, and upsilon production, and, second, the
initial energy densities can be more reliably estimated based on
perturbatively calculable quark and gluon $E_T$-production
\cite{BM87,EKRT99}, see also the talk by Hammon at this conference, and, the 
approach of refs. \cite{McLERRAN}. Let me point out that the pQCD calculation 
has now been done consistently in next-to-leading-order pQCD \cite{TE99}. 
\end{itemize}

In this conference, the detector overview and physics capabilities of
ALICE were reviewed by Morsch (ALICE).  The main goal of ALICE is to
study the behaviour of strongly interacting matter through observing
various independent QGP-signatures simultaneously. A recent review of
the physics at ALICE can be found in e.g. \cite{PBMqm99}. Let me, however,  
list some examples below:




\begin{itemize}

\item The global observables, such as charged particle multiplicity at
central rapidities, $dN_{ch}/dy$, and global transverse energy
distributions $dE_T/dy$ ($d\sigma/dE_T$) will be affected by the
initially produced densities (initial transverse energy is not
an observable unlike the final $E_T$) and subsequent $pdV$ work done
by the collective QGP-HG system against expansion \cite{EKRT99}.

\item Event-by-event fluctuations will be measured in many observables, 
such as forward-energy, charge particle multiplicity, HBT-correlations,
single event $\langle p_T\rangle$, particle ratios, $N_{ch}/N_{\gamma}$. 
In this way ALICE is expected to find the possible exotic events.

\item The transverse momentum spectra and particle abundancies will be
measured for all possible particle species ($\pi,\eta,\omega,\phi,
p,K,\Lambda,\Xi,\Omega,D,d,t,\alpha,...$). These will reflect the flow
effects and the freeze-out as discussed previously.

\item Identical particle HBT-correlations will be measured to give
information of the dynamical evolution, size and lifetime and
phase-time structure of the freeze-out process.

\item Open charm and open beauty measurements will provide a normalization for 
$J/\Psi$- and $\Upsilon$-suppression. As hard probes, $c\bar c$ and $b\bar b$
are produced very early on in the $AA$ collision. Consequently, they
will be very sensitive to the early and hot QCD phase of matter.

\item Energy-loss of high-$p_T$ partons in QGP \cite{BDMPS} can be studied 
via jet (and leading particle) measurements at ALICE. In this conference, 
the recent results on finite systems were discussed by Levai \cite{LEVAI}.

\item Muons at $2<\eta_{\mu}<4$ will be measured and $\mu^+\mu^-$ will
be studied. Especially interesting will be the dimuons coming from the
decays of $J/\Psi$ and $\Upsilon$ (if there are any $J/\Psi,
\Upsilon$ to decay!), as these will be directly probing the
deconfinement. Also intermediate mass muons will be very interesting,
due to the contributions from several sources, such as the Drell-Yan
process, decays of $D$- and $B$-mesons and thermal emission \cite{CHARM}.  
If $e^+e^-$ pairs will also be identified, as lately suggested,
modifications of the $\rho,\omega,\phi$-meson properties related to
the low-mass behaviour of $e^+e^-$ can be studied in detail.

\item Hard photons will also be an important tool to study the QGP.
Especially, in order to see any window for thermally radiated photons,
the background of direct photons and decay photons needs to be fully 
understood. As discussed for RHIC \cite{XNW}, the direct photon-jet 
events can possibly also be used to study energy loss of partons in QGP. 
In these calculations, and in calculations of any hard probes of QGP,
information of nuclear parton distributions and their scale
evolution \cite{EKS98} are needed.  The data from p+$A$ collisions 
will also be used to constrain the nuclear parton distributions. 

\end{itemize}

To compare some global characteristics of URHIC at SPS, RHIC and
LHC/ALICE, I have prepared the table 1. I would especially
like to draw your attention to  the expected long lifetime of the {\em
QGP-phase} at the LHC which is of the same order of magnitude as that
of the {\em whole} system at the SPS. Also, even though the system will not
freeze out simultaneously everywhere in space (roughly simultaneous
decoupling may happen in the transverse direction, except for the very
edges, but longitudinally there will be strong time dilation effects
at the LHC), I have defined a comoving volume by counting roughly how
many units of space-time rapidity are available along a proper
time curve. Notice how dramatic the increase in the ``size'' is when going
from the SPS to the LHC: the rest frame volume of a lead nucleus is
$V_{\rm Pb}=1150$ fm$^{3}$, so $400V_{\rm Pb}\sim 5\cdot10^{5}$
fm$^3$, and compared with the nuclear size parameters, one is talking about 
almost ``macroscopic'' quantities of strongly interacting matter at the LHC!

\begin{table} 
\begin{center} 
\begin{tabular}{llll}  
\br 
& {\bf SPS} & {\bf RHIC} &{\bf LHC}\\  
\mr 
$\sqrt s/A$ (GeV)& 17 	& 200 	& 5500\\
&&&\\
$\Delta Y$	 & 6	& 11	& 17 \\
&&&\\
$dN_{ch}/dy$	 & 400 	& 700-1500 & 3000-8000 \\
&&&\\
$\tau_{\rm tr}^{\rm PbPb}$ {(fm$/c$)}
& 1 & 0.1   & 0.005 \\
&&&\\
$\tau_0^{\rm QGP}$ {(fm$/c$)}
  & 1	& $\sim 0.2$ & 0.1 \\
&&&\\
$\varepsilon(\tau_0)$  $(\rm GeV/{\rm fm}^{3})$
& 3 	& 60	& 1000 \\
&&&\\
$\tau_{\rm QGP}$ (fm$/c$)
& $\lsim 2$ & 2-4	& $\gsim 10$  \\
&&&\\
$\tau_{\rm fo}$ {$({\rm fm}/c)$}
&$\sim 10$& 20-30 & 30-40\\
&&&\\
$V_{\rm com}(\tau_{fr})$&$8V_{\rm Pb}$& $90V_{\rm Pb}$ & $400V_{\rm Pb}$ \\
\br 
\end{tabular} 
\caption{ Some characteristics of the heaviest systems produced in
$AA$ at SPS, RHIC and LHC ($A\sim 200$). From the top: cms-energy
$\sqrt s$, available total rapidity interval $\Delta Y$, charged
particle multiplicity $dN_{ch}/dy$ at $y=0$, the transit time
$\tau_{\rm tr}$ of the Lorentz-contracted nuclei, the formation time
$\tau_0^{\rm QGP}$ of the QGP (lower limit) \cite{EKRT99}, the initial
energy density $\varepsilon(\tau_0)$ at the time of formation of the
QGP \cite{BJORKEN,EKRT99}, lifetime of the QGP phase $\tau{\rm QGP}$
\cite{EKRT99}, freeze-out time $\tau_{\rm fo}$ at $z=0$
\cite{TRANSVERSE}, ``comoving'' volume $V_{\rm com}(\tau_{fr})$ along
the freeze-out surface. $V_{\rm Pb}$ denotes the rest frame size of a
lead nucleus.  The value $dN_{ch}/dy=400$ is a measured one
\cite{JONESNA49}.  }
\end{center} 
\label{table1}
\end{table} 
  
To conclude the LHC-section for URHIC, let me also mention that 
the possibility of measuring  $J/\Psi$, $\Upsilon$
and $Z_0$ production in Pb+Pb collisions in the CMS detector 
at the LHC is being discussed, see e.g. \cite{PBMqm99}.

\section{My conclusions}

\subsection{SPS}

Analyses of several different observables from independent
measurements at the SPS have shown that a picture of production of
collectively behaving strongly interacting matter with large volume
and finite lifetime is now finally emerging. The systems produced in
collisions of heaviest nuclei (Pb+Pb, Pb+Au) are clearly different
from the ones produced in nucleon-nucleon and proton-nucleus
collisions.

Evidence of collective behaviour of matter is obtained from the
measured hadron yields and momentum spectra (NA44, NA49, WA97, WA98,
NA50, NA52): the systematic broadening of transverse momentum spectra
(sec.~2.1) is compatible with a behaviour typical for a system with
pressure and transverse flow.  The observation of elliptic and
directed flow (sec.~2.2) also strongly points to this direction (NA49,
WA98, NA52). The thermal model picture of the system reproduces the
measured particle ratios surprisingly well (sec.~2.3), and, the HBT
analysis of correlations in the momentum spectra (NA44, NA49, WA98) of
identical particles indicate that the system indeed is an extended
one, with a large volume and finite lifetime (sec.~2.4). The pattern
of the low-mass $e^+e^-$ enhancement, observed by CERES in S+Au and
Pb+Au collisions (sec.~2.7), also strongly lends support to the
picture of a collective, thermal, system of hadrons. The observed
various pieces of evidence of collectivity alone do not necessarily
imply that the system would have reached the QGP phase. The estimated
initial energy densities are sufficient (table~1) but the flow
studies and the event-by-event measurements of global variables
(sec.~2.5) have not shown any dynamical non-trivial fluctuations
typical for exotic events.

However, the fact that definite additional anomalies are observed 
gives more support to the QGP interpretation:

Strangeness production in $A$+$B$ (sec.~2.6) is globally enhanced
(NA35/49, WA97, NA38/50). Moreover, as observed by WA97, there is a
clear specific enhancement of multistrange hadrons: the yield of
$\Omega^-+\Omega^+$ per negative hadrons is enhanced by a factor $\sim
10$ in Pb+Pb collisions relative to p+p and p+Be. Microscopic
hadronic rescattering models imply that as multistrange hadrons are
difficult to produce due to the high mass-thresholds, the strangeness
increase must have an origin at the partonic level, before
hadronization.  The agreement of the thermal models (sec.~2.3) with
the measured hadron abundancies suggests that it is very likely that
the system is indeed thermalized already before the completion of 
hadronization.

Finally, the best candidate for a direct QGP signature so far is the
anomalous suppression of $J/\Psi$ observed in central Pb+Pb collisions
by NA50 (sec.~2.8).  It is very difficult to explain the additional,
anomalous, suppression of $J/\Psi$ (i.e. the suppression in addition
to the expected suppression in nuclear matter) without invoking
partonic degrees of freedom and deconfined matter.  I would like to
emphasize that in any possible alternative scenario, {\em all} the
previously mentioned features of a collective system and the
importance of partonic degrees of freedom will also have to be
explained {\em simultaneously}. To my best knowledge, such an
alternative interpretation does not exist so far.

My conclusion of the results from URHIC at the SPS therefore is that
collective behaviour of matter is observed but so far there is not yet 
a 100\% certain proof of an observation of the QGP, although the
strangeness enhancements and the anomalous $J/\Psi$
suppression strongly suggest it.

\subsection{SPS $\rightarrow$ RHIC $\rightarrow$ LHC/ALICE}

In increasing the cms-energy of $AA$-collisions, the quantitative gains
are obvious (sec.~3.2): the system becomes initially denser and
hotter, it forms faster and stays collective for a longer time,
i.e. its lifetime grows and volume increases. The longer the QGP-phase
itself can be made, the better chances there are to observe direct QGP
signatures.

The higher the cms-energy is, the higher are also the rates for the truly
{\em hard probes} of QGP which cannot be studied at the SPS. At RHIC
and LHC/ALICE, these additional QGP probes include $\Upsilon$
(especially its suppression) and open beauty production (at the LHC in
particular), jets, high-$p_T$ direct photons, and high mass dileptons.
Measurements of $Z^0$ may also be possible in Pb+Pb collisions at
the CMS.  At the high energies of RHIC and especially at the
LHC/ALICE, the hot initial conditions of the QGP are also expected to
become more reliably calculable based on perturbative QCD.

According to the official plans at the time of this meeting, RHIC at
BNL should be in operation in December 1999, so the first exciting data 
on global variables at high energies will be available in year 2000. 
A bit further in the future, LHC/ALICE at CERN
will start in 2005, completing the extensive experimental program of
ultrarelativistic heavy ion collisions. As the URHIC program at the
SPS is currently teaching us, at RHIC and ALICE the observation of the QGP will
be confirmed by studying carefully {\em several independent observables
simultaneously}.  After the results from LHC/ALICE, a broad
range of the QCD phase diagram has been experimentally probed, and
more light has been shed to the theory of elementary particle condensed 
matter physics and to the early evolution of  our Universe.
\vspace{0.1cm}

{\bf Acknowledgements.}  I thank K. Kajantie, U. Heinz, V. Ruuskanen,
J. Schukraft, C. Louren{\c c}o, P. Huovinen, V. Kolhinen and K. Tuominen
for discussions and help in preparing this talk.

\end{document}